\documentclass[prl,aps,twocolumn,showpacs,preprintnumbers,amsmath,amssymb,twoside]{revtex4-1}

\usepackage{setspace}
\usepackage{times}
\usepackage{graphicx}
\usepackage{pstricks}
\usepackage{dcolumn}		
\usepackage{bm}			
\usepackage{ulem}
\usepackage{units}		
\usepackage{hyperref}
\usepackage{color}
\hypersetup{colorlinks=true,urlcolor=blue,linkcolor=blue,citecolor=blue}

\newcommand{\cco}{CaCu$_2$O$_3$}
\newcommand{\sco}{Sr$_2$CuO$_3$}
\newcommand{\cuo}{CuO$_4$}
\newcommand{\q}{${\bf q}$}
\newcommand{\cul}{Cu $L_3$ edge}
\newcommand{\clr}{Cu $L_3$ RIXS}

\newcommand{\jg}[1]{\textcolor{black}{#1}}
\newcommand{\vb}[1]{\textcolor{black}{#1}}
\newcommand{\jb}[1]{\textcolor{black}{#1}}

\newcommand{\kwf}[1]{\textcolor{black}{#1}}
\newcommand{\kw}[1]{\textcolor{black}{#1}}
\newcommand{\vbf}[1]{\textcolor{black}{#1}}

\newcommand{\vbb}[1]{\textcolor{black}{#1}}
\newcommand{\vbbb}[1]{\textcolor{black}{#1}}
\newcommand{\kwn}[1]{\textcolor{black}{#1}}
\newcommand{\kwnn}[1]{\textcolor{black}{#1}} 
\newcommand{\vbn}[1]{\textcolor{black}{#1}} 

\begin{document}

\title{Orbital control of effective dimensionality: from spin-orbital fractionalization to confinement in the anisotropic ladder system \cco}

\author{Valentina Bisogni$^{1,2,3}$}
\altaffiliation[Present address: ]{National Synchrotron Light Source II, Brookhaven National Laboratory, Upton, NY 11973-5000, USA}
\email{bisogni@bnl.gov}
\author{Krzysztof Wohlfeld$^{1,4,5}$}
\altaffiliation[Present address: ]{Stanford Institute for Materials and Energy Sciences, SLAC National Laboratory and Stanford University, 2575 Sand Hill Road, Menlo Park, CA 94025, USA
and Institute of Theoretical Physics, Faculty of Physics, University of Warsaw, Pasteura 5, PL-02093 Warsaw, Poland}
\email{wohlfeld@stanford.edu}
\author{Satoshi Nishimoto$^{1}$}
\author{Claude Monney$^{2,6}$}
\altaffiliation[Present address: ]{Department of Physics, University of Z\"{u}rich, Winterthurerstrasse 190, CH-8057 Z\"{u}rich, Switzerland}
\author{Jan Trinckauf$^{1}$}
\author{Kejin Zhou$^{2,7}$}
\altaffiliation[Present address: ]{Diamond Light Source, Harwell Science and Innovation Campus, Didcot, Oxfordshire OX11 0DE, UK}
\author{Roberto Kraus$^{1}$}
\author{Klaus Koepernik$^{1}$}
\author{Chinnathambi Sekar$^{1,8}$}
\altaffiliation[Present address: ]{Department of Bioelectronics and Biosensors, Alagappa University, Karaikudi-630 003, Tamilnadu, India}
\author{Vladimir Strocov$^{2}$}
\author{Bernd B\"uchner$^{1,9}$}
\author{Thorsten Schmitt$^{2}$}
\author{Jeroen van den Brink$^{1,9}$}
\author{Jochen Geck$^{1}$}

\affiliation{$^{1}$IFW Dresden, Helmholtzstrasse 20, 01069 Dresden, Germany\\
$^{2}$Swiss Light Source, Paul Scherrer Insitute, CH-5232  Villigen PSI, Switzerland\\
$^{3}$National Synchrotron Light Source II, Brookhaven National Laboratory, Upton, NY 11973-5000, USA\\
$^{4}$Stanford Institute for Materials and Energy Sciences, SLAC National Laboratory and Stanford University, 2575 Sand Hill Road, Menlo Park, CA 94025, USA\\
$^{5}$Institute of Theoretical Physics, Faculty of Physics, University of Warsaw, Pasteura 5, PL-02093 Warsaw, Poland\\
$^{6}$Department of Physics, University of Z\"{u}rich, Winterthurerstrasse 190, CH-8057 Z\"{u}rich, Switzerland\\
$^{7}$Diamond Light Source, Harwell Science and Innovation Campus,Didcot, Oxfordshire OX11 0DE, UK\\
$^{8}$Department of Bioelectronics and Biosensors, Alagappa University, Karaikudi-630 003, Tamilnadu, India\\
$^{9}$Department of Physics, Technical University Dresden, D-1062 Dresden, Germany\\}

\date{Received: \today}

\begin{abstract}
Fractionalization of an electronic quasiparticle into spin, charge and orbital parts is a fundamental and characteristic property of interacting electrons in one dimension.
\kwnn{However, real materials are never strictly one-dimensional and the fractionalization phenomena are hard to observe.}
\kwnn{Here}
we studied the spin and orbital excitations of the anisotropic ladder material \cco, 
\kwnn{whose electronic structure is not one-dimensional.}
Combining high-resolution resonant inelastic x-ray scattering experiments with theoretical model calculations we show that: (i) spin-orbital fractionalization occurs in \cco~along the leg direction {\it x} through the {\it xz} orbital channel as in a 1D system; 
 and (ii) no fractionalization is observed for the {\it xy} orbital, which extends in both leg and rung direction, contrary to a 1D system. 
We conclude that the directional character of the orbital hopping can select different degrees of dimensionality.
\vbn{Using additional model calculations}, we show that spin-orbital separation is {\it generally} far more robust than the spin-charge separation. This is not only due to the already mentioned selection realized by the orbital hopping, but also due to the fact that spinons are faster than the orbitons.

\end{abstract}


\maketitle

An electron in a Mott insulator is characterized by \vbn{fundamental quantum numbers, representing its} charge, its spin and the orbital it occupies. But when electrons are confined to one dimension, these basic properties of the electron can break apart (fractionalize), forming three independently propagating excitations called spinon, holon and orbiton. While this very fundamental property of one-dimensional systems is both experimentally and theoretically well-established \cite{Voit1995, Kim1996, Wohlfeld2011, Schlappa2012, Moreno2013}, its relevance in higher dimensions is heavily debated, in particular in relation to the high temperature superconductivity in two-dimensional cuprates \cite{Anderson1987,Senthil2001,Holt2012}.

\begin{figure}[t!]
\center{
\includegraphics[width=0.93\columnwidth]{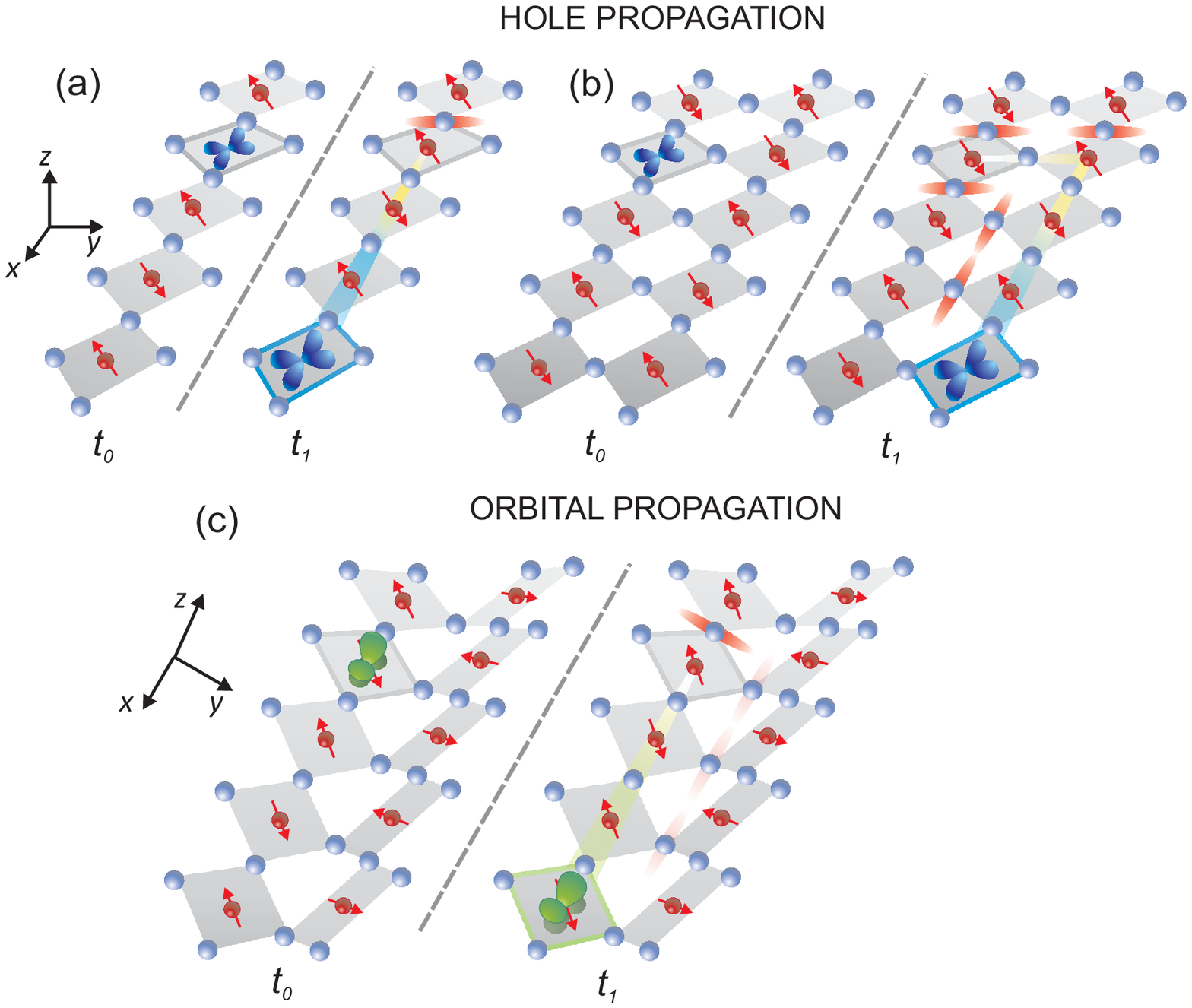}}
\caption{(Color online) \kw{Schematic view of the} hole/orbiton propagation in an $S$=1/2 environment. (a) A hole with 3$d_{x^2-y^2}$ symmetry moves from its original position (second plaquette from top, left chain) \jb{at time $t_0$ } and \jb{while propagating} \jb{at a later time $t_1$} it has created only one magnetic domain wall at the first hop (cigar shape in the right chain). (b) In a 2D lattice, or equivalently in a two-leg ladder, the \vb{moving} 3$d_{x^2-y^2}$ hole creates a trail of 
magnetic domain walls. (c) In the buckled two-leg ladder \cco~ studied here, when the 3$d_{xz}$ {\it orbiton} moves (second plaquette from top, left chain), it creates just one magnetic domain wall as in (a) not only because interleg domain walls (lightly shaded cigar shape) can be neglected (due to very weak spin interaction along the rung) but also because it can move {\it solely} along one of the legs of the ladders \vb{(due to the directional hopping of the 3$d_{xz}$ orbital).}}
\label{fig:cartoon}
\end{figure}

The basic conceptual difference between \vbbb{one-dimensional} (1D) and \kwn{higher dimensional systems, such as ladders or} \vbb{two-dimensional materials} (2D), can be illustrated by taking away an electron, having spin and charge, from either an antiferromagnetic (AF) chain or an AF plane [see Fig. \ref{fig:cartoon}(a) and (b)], and \jb{subsequently} consider the propagation of the resulting hole~\cite{nhole} \jb{in these two different situations}. In the 1D case, the hole can start propagating freely after exciting only one spinon (a magnetic domain wall in the chain) and can consequently separate into a holon and a spinon [Fig.\,\ref{fig:cartoon}(a)] \cite{Voit1995}. But when hopping between sites in \vb{an} AF plane \kwn{or ladder}, the hole leaves behind a long trail of wrongly aligned spins [Fig.\,\ref{fig:cartoon}(b)] \cite{Martinez1991, Troyer1996, White1997}. This trail acts as a string-potential that tends to confine the hole to its starting position and thereby to bind the spin and charge of the hole~\kwn{\cite{Martinez1991, Troyer1996, White1997, nladder}}. 

These basic examples illustrate that whereas spin-charge separation is a hallmark of 1D systems, any system beyond strictly 1D is much more complex because of the 
coupling \jg{of the charge carriers} to the inherently strongly fluctuating quantum spin-background \cite{Voit1995, Martinez1991}. It actually remains to be established if and how any of the peculiar 1D physics carries over to higher dimensions \cite{Dagotto1996}. In this paper, we take a 
pragmatic approach to this problem: we try to verify whether a realistic system that is not strictly 1D -- and thus for example spin-charge separation is hard to observe -- might nevertheless show fractionalization phenomena for some quantum numbers, because of the reduced effective dimensionality of these `specific' quantum numbers.

In this Letter, we present a combined experimental and theoretical study of the AF anisotropic spin-ladder system CaCu$_2$O$_3$,
cf. Figure \ref{fig:cartoon}(c). We look for signatures of spin-orbital separation, which is conceptually analogous to the spin-charge separation described above \cite{Wohlfeld2011,Schlappa2012}. To this purpose, we use the 
resonant inelastic X-ray scattering (RIXS) method that is inherently sensitive to \jb{both} spin and orbital excitations in cuprate materials\,\cite{Ghiringhelli2004,Braicovich2009,Schlappa2009,Ament2011,Schlappa2012}.  Our experimental results together with theoretical model calculations show that \vbn{spin-orbital fractionalization occurs in \cco~ for the {\it xz} orbital channel -- as in an ideal 1D system -- while confinement is found for the {\it xy} orbital channel. From this result, we convey that diverse orbital symmetries select different degrees of dimensionality in the same system.}
Moreover, we demonstrate that spin-orbital separation is in general much more robust than spin-charge separation.

\begin{figure}[t!]
\center{
\includegraphics[width=0.95\columnwidth]{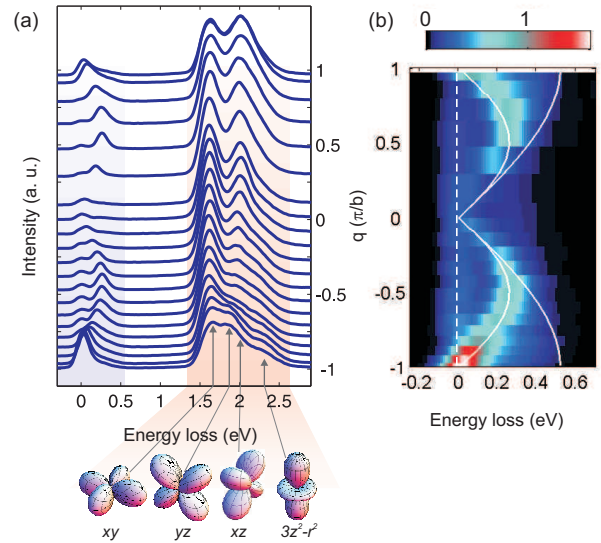}}
\caption{(Color Online) RIXS spectra of \cco~ measured \vb{at the maximum of the Cu $L_3$ resonance}, for in-plane polarization of the incident x-rays, at 40 K. (a) The spectra \vb{in the vertical cascade} correspond to different momenta \q, \vb{transferred along the leg direction $x$ ($b \simeq 4.1$ \AA)}. At low energies (below 0.3 eV), the spinon dispersion clearly emerges \cite{Bisogni2013}. At high energies, orbital excitations dominate the spectra. The arrows identify the \vb{corresponding} orbital symmetry of the hole in the RIXS final state, with the following energy splittings: 3$d_{x^2-y^2}$ at 0 eV (ground state), 3$d_{xy}$ at 1.62 eV, 3$d_{yz}$ at 1.84 eV, 3$d_{xz}$ at 2.03 eV and 3$d_{3z^2-r^2}$ at 2.35 eV (see supp. mat. \cite{SM} and Tab.\,S1). (b) RIXS intensity map of the dispersing magnetic excitation. The solid white lines represent the lower and the upper boundaries of the multi-spinon continuum of a $S$=1/2 Heisenberg chain as in \cite{Lake2010}. The dashed white line marks the zero energy loss.}
\label{fig:exp}
\end{figure}

\kw{\cco \ is a buckled two-leg spin ladder system composed of corner-sharing \cuo~ plaquettes [Fig.\,\ref{fig:cartoon}(c)] \cite{Bordas2005, Lake2010}} \vb{building a system of coupled spin chains \kwf{parallel to the ladder leg direction $x$} with a Cu-O-Cu angle of 123$^\circ$ along the \kwf{ladder} rung direction $y$.}
Just as the well-known 2D cuprate compounds, the ground state configuration of this system is dominated by Cu$^{2+}$ sites in each plaquette with \kw{$S$}=1/2 and one hole~\cite{nhole} \kw{localized} in the $3d_{x^2-y^2}$ orbital. 

\jg{In order to investigate the spin and orbital dynamics of \cco, we performed high-resolution RIXS experiments at the Cu \kw{$L$}$_3$ edge \vbbb{(for more details, cf. Ref. \cite{SM})}. RIXS enables to map out the dispersion of these excitations across the first Brillouin zone of \cco. The experiments were performed using the state-of-the-art SAXES spectrometer \cite{saxes} installed at the ADRESS beamline of the Swiss Light Source at the Paul Scherre Institute\,\cite{Strocov2010}.}

\jb{The RIXS spectra, shown in Figure \ref{fig:exp}(a) as a vertical cascade for increasing momentum \q~ transferred along the leg direction $x$, reveal the presence of strongly dispersing magnetic excitations at energies between 0\,eV and 0.5\,eV.}
\jg{\jb{As indicated in Figure \ref{fig:exp}(b), their} momentum dependence exactly tracks 
the well-known two-spinon dispersion \cite{Bisogni2013}, which is in perfect agreement with 
previous neutron scattering \jb{data} \,\cite{Lake2010} and confirms \kw{the prevalent 1D nature of the {\it spin} system. It is important to point out however that the low-dimensional magnetism of \cco~ is not reflected in the electronic band structure, which exhibits a pronounced 2D character \cite{Bordas2005}.}}

In the energy region between 1.5 eV to 2.6 eV we observe orbital excitations, 
\jb{corresponding to} the hole\kw{~\cite{nhole}} in the 3$d_{x^2-y^2}$ ground state \jb{being} \vb{excited} into a different orbital. We can unambiguously identify these excitations and their energies by combining state-of-the-art quantum chemical calculations \cite{Huang2011}, which provide realistic estimates for the orbital sequence and the energy splittings, with local calculations of the RIXS cross-section \kwf{based on single ion model (also referred to as `local model' in what follows)}\,\cite{Moretti2011}, cf. Figure \ref{fig:exp} and \cite{SM}.

\begin{figure*}[t!]
\center{
\includegraphics[width=1.9\columnwidth]{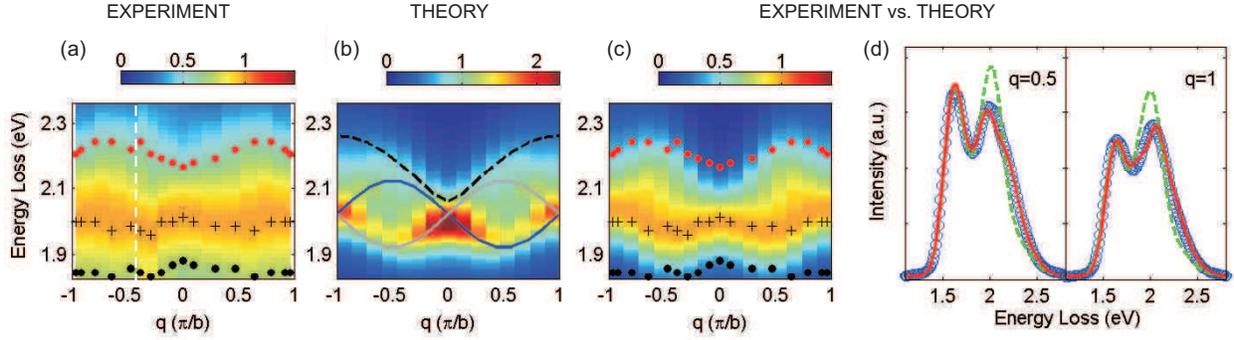}}
\caption{(Color Online) Experimental and theoretical 3$d_{xz}$ \vb{orbital} dispersion. (a) Color map of the 3$d_{xz}$ excitation lineshape.  Each RIXS spectrum, in the range 1.8-2.35 eV, is normalized to the 3$d_{xz}$ orbital excitation maximum. Data between $-1<\q<-0.6$ are masked by the close 3$d_{yz}$ excitation, which is strong at small \q, and have thus been replaced by the mirrored data from the positive \q \/ range. The red (top) and the black (bottom) dots define the 50\% and 35\% intensity drop with respect to the peak position. The latter is marked by black crosses. (b) Color map of the exact diagonalization solution for the spin-orbital separation model of \cco \/ [see Eq.\,(\ref{eq:tJ1})]. The lines follow from the orbital-spin separation Ansatz \cite{Schlappa2012} applied to \cco~ which shows the pure orbiton dispersion (solid line) and the edge (dashed line) of the spinon-orbiton continuum. (c) Normalized theoretical map (b), after broadening by experimental resolution ($\Delta$E=130 meV). Dots are taken from the experimental map (a). (d) Comparison between RIXS experimental lineshapes (blue open dots), local model \kw{for all orbital excitations} (dashed green), and combination of local model (\kw{for} 3$d_{xy}$, 3$d_{yz}$ and 3$d_{3z^2-r^2}$) and spin-orbital separation model for 3$d_{xz}$ (red solid line) \cite{SM}.}
\label{fig:comparison}
\end{figure*}

The good agreement of this local model and the experimental data (see suppl. mat. \cite{SM}) implies that the  orbital excitations into the 3$d_{xy}$, 3$d_{yz}$, and 3$d_{3z^2-r^2}$ are in fact local excitations within a CuO$_4$ plaquette. However, \jb{we observe that} in case of  the 3$d_{xz}$ orbital excitation such local description fails \cite{SM} and a propagating {\it orbiton} is observed, see Figure \ref{fig:comparison}(a): by normalizing the spectra for each \q \/ to its maximum value, a clear momentum dependence is revealed, which consists of a peak shift of 50 meV towards lower energies together with a strong increase in width by almost a factor of 2 for increasing \q. Such momentum \vb{dispersion} is unusual and is in fact \vb{a}  \jb{direct} indication of spin-orbital separation~\kw{\cite{Wohlfeld2011, Schlappa2012}}. 

In order to verify that spin-orbital separation indeed takes place in \cco, we analyzed the spin-orbital dynamics of this material in terms of an effective $t$--$J$ model. The various hopping integrals \jb{in the}
model are determined using density functional theory calculations \cite{Koepernik1999}. Based on these parameters, the effective $t$--$J$ model ~\cite{Wohlfeld2011} for \cco\/ can be constructed, which takes into account all possible superexchange processes with virtual states both on copper and on oxygen sites \cite{Wohlfeld2013} [see \vbn{Fig.~S2(a)  in \cite{SM} for details}]:
\begin{align} \label{eq:tJ1}
{\mathcal {{H}}} =&   - J^{xz}_{\rm {leg}} \!\sum_{\langle {\bf i}, {\bf j} \rangle || x , \sigma} 
({o}^\dag_{{\bf i} \sigma, {xz}}
{o}_{{\bf j} \sigma, {xz}}\! +\! h.c.) 
\kwf{+E_0^{xz} \sum_{\bf i} n_{{\bf i}, xz}}
\nonumber \\ 
 &+ J_{\rm {rung}} \sum_{\langle {\bf i}, {\bf j} \rangle || y } {\bf S}_{{\bf i} } \cdot {\bf S}_{\bf j} 
 + J_{\rm {leg}} \sum_{\langle {\bf i}, {\bf j} \rangle || x } {\bf S}_{{\bf i} } \cdot {\bf S}_{{\bf j}},
\end{align}

Here ${o}^\dag_{{\bf i} \sigma}$ creates a hole in the excited 3$d_{xz}$ orbital on site ${\bf i}$ with spin $\sigma$, \kwf{$n_{{\bf i}, xz}$ counts the number of holes in the 3$d_{xz}$ orbital on site ${\bf i}$,} and ${\bf S}_{\bf i}$ is a spin $S=1/2$ operator on site ${\bf i}$. The values of the parameters are (for further details see suppl. mat. \cite{SM}): the on-site energy of the 3$d_{xz}$ orbital excitation \kwf{$E_0^{xz}=\vbf{2.03}$ eV}, $J^{xz}_{\rm {leg}} \simeq 51$ meV for the 3$d_{xz}$ orbital superexchange, $J_{\rm {leg}} \simeq 134$ meV, and $J_{\rm {rung}}\simeq 11$ meV for the spin superexchange along the leg and along the rung, respectively. We underline that, in the present case, \vbbb{the orbital part of the Hamiltonian has by itself a 1D character
: the reason behind this lies in the} intrinsic \vbn{one-}dimensionality of the $3d_{xz}$ orbital hopping, which projects solely along {\it x} (within the ladder plane) and is completely unaffected by the neighboring ladders [refer to Fig.~S2(a) in \cite{SM}]. 

Using Lanczos exact diagonalization, we \jb{determined} the orbiton spectral function \cite{Wohlfeld2011} for a single 3$d_{xz}$ orbital excitation introduced into the AF ladder, which propagates via the Hamiltonian ${\mathcal {{H}}}$ on a $14\times2$-site ladder lattice. The obtained spectral function is shown in Figure \ref{fig:comparison}(b). \kw{In order to facilitate the comparison to the experiment, we normalize the theoretical spectra from Figure \ref{fig:comparison}(b) in the same manner as the experimental data and, after including the instrumental broadening, we present the results in Figure \ref{fig:comparison}(c).} Excellent agreement between theory and experiment is observed. This is further confirmed by comparing calculated and measured RIXS responses at two fixed \q-values [see Fig.\,\ref{fig:comparison}(d) \kw{and Ref.} \cite{SM}].

The comparison between the \kw{theoretical} spectral function in Figure \ref{fig:comparison}(b) with the one obtained for the purely 1D case~\cite{Wohlfeld2011} reveals that the 3$d_{xz}$ excitation in \cco\/ exhibits all typical features of spin-orbital separation. This observation is also verified by a detailed analysis of Equation (\ref{eq:tJ1}), which at energy scales $E\gg J_{\rm {rung}}$ effectively describes a 1D $t$--$J$ model \vbb{giving qualitatively the same spectral response as for the 1D case}:  having a 1D spin superexchange in the relevant energy scale is in fact a necessary condition for observing fractionalization. \jb{Higher dimensional effects become important only at energy scales $E \ll J_{\rm {rung}}$ \cite{Lake2010}, which are presently not accessible by the experimentally available RIXS energy resolution.} 
\begin{figure}[t!]
\center{
\includegraphics[width=0.85\columnwidth]{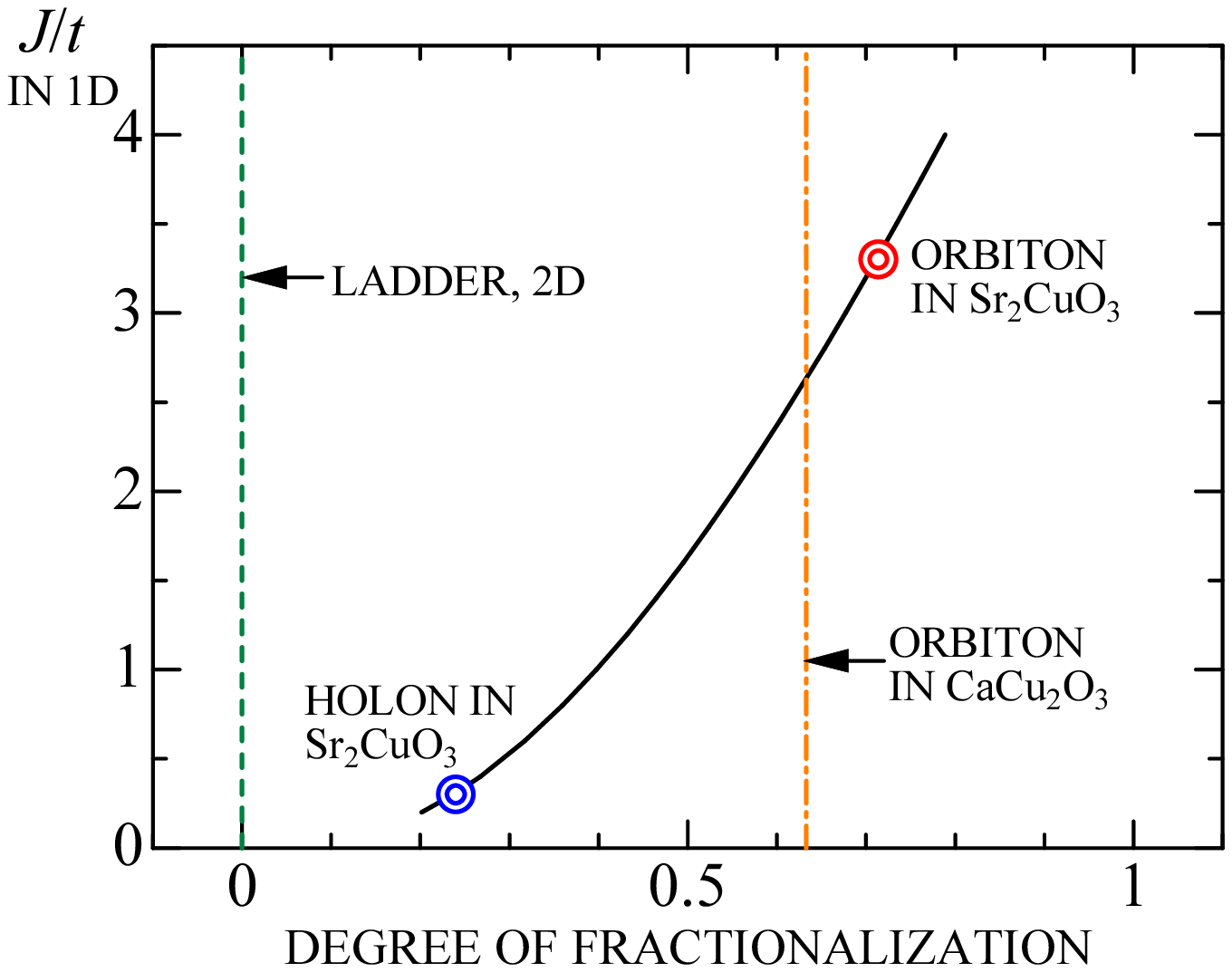}}
 \caption{DOF in various effective $t$--$J$ models as obtained using the finite size scaled exact diagonalization: for the isotropic ladder $t$--$J$ model (dashed line, DOF=0), for the anisotropic ladder $t$--$J$ model Eq. (1) describing the spin-orbital separation in \cco~ (dot dashed line, DOF$\simeq$0.62). These cases are compared with the ideal 1D $t$--$J$ model, for which the DOF is calculated as a function of $J/t$ (solid line): $J/t \simeq 0.4$ ($J/t\simeq 3.32$) describes the spin-charge (spin-orbital) separation in \sco.}
\label{fig:fractionalization}
\end{figure}

In a pure 1D system, however, spin-orbital separation manifests also for the 3$d_{xy}$ orbital \cite{Schlappa2012}. In the present case instead, the good agreement between the 3$d_{xy}$ orbital excitation and the local model [see Fig.~\ref{fig:comparison}(d) at around 1.6 eV and Ref. \cite{SM}] seems not to support fractionalization for this orbital channel.

To investigate the possibility of spin-orbital fractionalization for the $3d_{xy}$ orbital in \cco, a similar analysis to the $3d_{xz}$ case has been done. Two finite paths for the orbital superexchange are found in this case, respectively along the leg and the rung directions as the hole can hop both along {\it x} and {\it y}  [see Fig.~S2(b) and \cite{SM} for more details]. 
Nevertheless, it occurs that the hopping along the leg of the ladder is {\it blocked} due to a peculiar interplay of the Pauli principle and a strong inter-ladder hopping between the $2p_y$ oxygen orbital in the leg and the 3$d_{x^2-y^2}$ copper orbital in the neighboring ladder [cf. Fig. S2(b) in \cite{SM}]. Because of that, there is a nonzero probability of having the $2p_y$ oxygen orbital in the leg occupied by a hole coming from the neighboring ladder. Considering moreover 
that the spin of the {\it traveling} hole from the 3$d_{xy}$ orbital is randomly oriented with respect to the spin of the hole in the $2p_y$ orbital, the 3$d_{xy}$ hopping along {\it x} can be blocked in order not to violate the Pauli principle. In this way the coherent travel of the 3$d_{xy}$  orbital along the leg is suppressed: the 3d$_{xy}$ orbital excitation thus results in a localized excitation.

The above discussion of the $3d_{xy}$ case shows the importance of the directionality of the orbital motion in establishing the fractionalization phenomenon.
We observe spin-orbital separation for the strictly 1D 
3$d_{xz}$ orbital excitation, though, at the same time the separation is not
possible for the 
3$d_{xy}$ 
orbital excitation, which is strongly affected by inter-ladder couplings and, hence, behaves as in a 2D system. Note that the spin exchange is effectively 1D through both orbitals.

\vbn{Finally, to quantitatively compare the  spin-orbital separation observed in \cco\ and in other systems exhibiting spin-orbital or spin-charge separation, we consider here the ``degree of fractionalization'' (DOF), see Fig.~\ref{fig:fractionalization}: this index expresses how well the spinon and the orbiton or holon are separated from each other, and it ranges from 0 (not separated) to 1 (fully separated). Namely, given the correlation between the orbital (or charge) degree of freedom at site ${\bf i=1}$ and the spin at site ${\bf i} + {\bf r}$, the DOF is the ratio between the correlation for r=$\infty$ and r=1 (see  supp. mat. \cite{SM} for more details). Despite the presence of finite inter-leg interaction, we have verified that the spin-orbital separation in \cco~ is almost as strong as the one in \sco. Moreover, surprisingly, it is much stronger than the spin-charge separation in the strictly 1D system~ (Fig.~\ref{fig:fractionalization}). Besides the directional character of the orbital, the reason behind this behavior lies in the different ratio of the spinon / orbiton and spinon / holon velocities. In fact the spinon can move away much quicker from the orbiton than from the holon, i.e. from the quasiparticle at which we look in the experiment, allowing for an `easier' separation from the spinon in the spin-orbital case.}

\kwnn{In conclusion, {the present RIXS experiments and theoretical analysis thus demonstrate} spin-orbital separation for the 3$d_{xz}$ orbital in the {anisotropic} ladder material \cco. The spin-orbital separation is therefore not limited to ideal 1D systems only, but can also survive in systems with the electronic structure being not strictly 1D. This robustness of the fractionalization of the orbital excitation is related to {the incidence of} three features: (i) the spin excitations in the weakly coupled spin ladder \cco~ are essentially spinons {on the here relevant energy scale}, (ii) the motion of the orbital excitation {\it can be} 1D due to the typical directional character of orbital hoppings ({\it note} here the 3$d_{xy}$ orbiton case for which the hopping is not 1D and the fractionalization does not take place)~\cite{Harris2003, Wohlfeld2008}, and (iii) spinons are much faster than orbitons (typically $J_{\rm {leg}} \gg J^{xz}_{\rm {leg}}$, cf. \cite{Wohlfeld2011}) which {allows for an} easy separation of the two. While the first condition requires 1D spin exchange interactions and is similarly valid for the spin-charge separation phenomenon, the other two conditions are generic to the spin-orbital separation phenomenon only and are rather easy to achieve in a number of systems whose band structure is not 1D. 
{This therefore suggests} that the spin-orbital separation can be realized in many other correlated systems {that lack} spin-charge separation because they are {\it not} strictly \jb{enough} 1D.} 

{\it Acknowledgements --} This work was performed at the ADRESS beamline of the Swiss Light Source at the Paul Scherrer Institut, Switzerland. This project was supported by the Swiss National Science Foundation and its National Centre of Competence in Research MaNEP. This research has been funded by the German Science Foundation within the D-A-CH program (SNSF Research Grant 200021L\_141325 and Grant GE 1647/3-1). Further support has been provided by the Swiss National Science Foundation through the Sinergia network Mott Physics Beyond the Heisenberg Model (MPBH). V.\,B. acknowledges the financial support from Deutscher Akademischer Austausch Dienst \vbbb{and the European Community's Seventh Framework Programme (FP7/2007-2013) under grant agreement No.\,290605 (PSI-FELLOW/COFUND)}. K.\,W. acknowledges support from the Alexander von Humboldt Foundation, from the Polish National Science Center (NCN) under Project No.~2012/04/A/ST3/00331 and by the Department of Energy, Office of Basic Energy Sciences, Division of Materials Sciences and Engineering (DMSE) under Contract Nos. DE-AC02-76SF00515 (Stanford/SIMES). J.\,G., R.\,K. and V.\,B. gratefully acknowledge the financial support through the Emmy-Noether Program (Grant GE1647/2-1). We are very grateful for insightful discussions with M. Daghofer, S. Kourtis, and L. Hozoi.

\clearpage

\begin{center}
\bf{Supplementary Material}
\end{center}
\setcounter{page}{1}

\section{A. Methods}

We measured a \cco~ single crystal grown by the travelling floating zone method \cite{Sekar2005}. The compound is an anisotropic two-leg ladder system characterised by two parallel CuO$_3$ chains \kwf{along the ladder leg direction $x$}, forming an angle of 123$^\circ$ \kwf{along the ladder rung direction $y$} [see Fig.\,1(c) \kw{in the main text}]. This buckling strongly influences the electronic hopping integrals of the system, thus realizing a configuration intermediate between 1D and 2D (
\kwf{cf.} following Sec.\, {\bf C} and {\bf D}).  

The measurements were done at a temperature of 40 K (above the \kw{antiferromagnetic ordering} temperature $T_{N}$=25 K). The crystal with a [001] surface was cleaved \kwf{{\it in situ}} at a pressure of 5$\times 10^{-10}$ mbar and oriented \kw{along} the \kw{ladder} direction $x$, being parallel to the scattering plane. We investigated the system with RIXS at the \cul, since it is well-established that \clr~ can probe orbital excitations with high sensitivity \cite{Ghiringhelli2004,Moretti2011,Ament2011}, by exciting at the maximum of the resonance (E=931.5 eV). The SAXES spectrometer of the ADRESS beamline of the Swiss Light Source was prepared with a scattering angle of 130$^\circ$, achieving a total energy resolution of 130 meV. The RIXS spectra were acquired for different values of transferred momentum \q~ along the $x$ direction.  By changing the incident angle onto the sample surface from 5$^\circ$ to 125$^\circ$, a \q-range of almost -1 to 1 in units of $\pi/b$ ($b\simeq 4.1$ \AA) could be spanned. All the spectra shown in the paper and supplementary information were measured with linear polarization of the incident x-rays parallel to the scattering plane, referred to as $in-plane$ polarization. Each spectrum \kw{was} recorded for a total time of 30 minutes.

\section{B. Identifying orbital excitation in \cco}

At energy losses greater than 1.5 eV \kwf{[cf.} Fig.\,2(a) in the main text], four different peaks are distinguishable in the RIXS spectra of \cco~  and assigned to orbital excitations. In order to indentify the single contributions, we make use of two theoretical tools. As mentioned in the main text, the energy splittings of the orbital excitations are in good agreement with quantum chemical calculations \cite{Huang2011}, which provide a crucial hint for assigning the sequence of orbital excitation energies. However, the individual theoretical energy values, although very close to the experiment, are overall slightly underestimated. Therefore, in order to finely determine the energies of the orbital excitations, we made a proper decomposition of the experimental data, as shown later on in this section. Concerning the intensities, a qualitative agreement is obtained for the different scattering geometries  by using a single ion model to calculate the RIXS cross section (so-called local model, see main text) \cite{Moretti2011} [see Fig.\,S1(b) and (c)]. The individual theoretical responses of each orbital excitation calculated within this model are shown in Figure S1(a) for  $in-plane$ polarization, after having exactly included the present geometry of the \cuo~plaquette with respect to the scattering plane.

\begin{figure*}[t!]
\includegraphics[width=1.35\columnwidth]{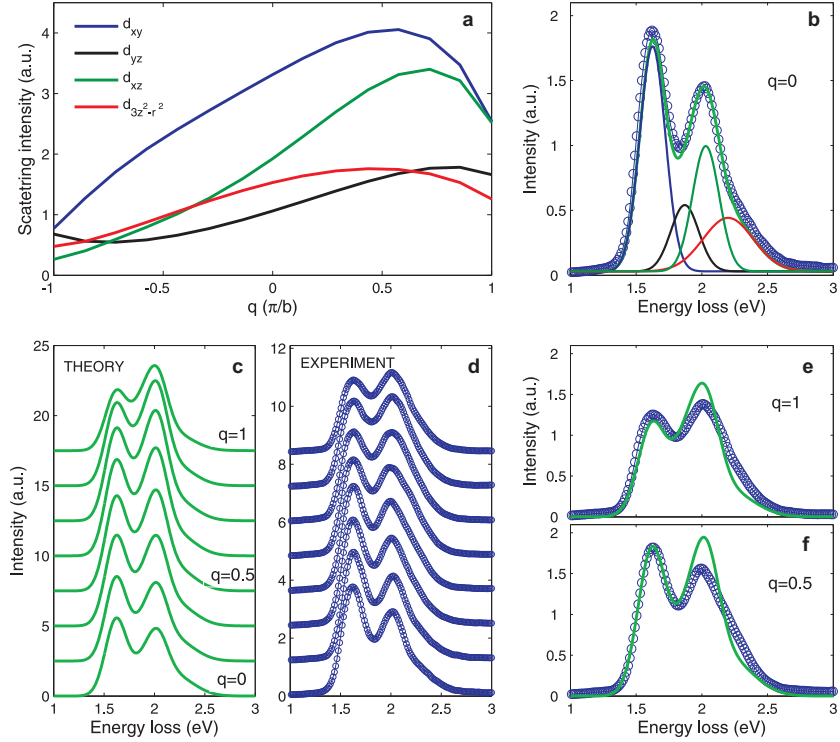}
\caption{RIXS orbital excitation of \cco~in 
\kwf{the local model.}
{\bf a}, Single ion responses of 3$d_{xy}$, 3$d_{yz}$, 3$d_{xz}$ and 3$d_{3z^2-r^2}$ excitations, calculated by considering an angle of 61.5$^\circ$ between the plaquette plane and the scattering plane (direct consequence of the buckled structure of \cco). {\bf b}, Theoretical RIXS spectrum (green solid line) constructed as the sum of the single Gaussian contributions corresponding to the different $dd$ excitations [thin solid lines, with the same color code as in {\bf a}]. The blue dotted line is the corresponding experimental spectrum at \q=0. All the theoretical spectra in this figure are calculated using the Gaussian parameters of Tab.\,\ref{t1}. {\bf c} ({\bf d}), Vertical stack of theoretical (experimental) RIXS spectra, for positive \q~ values (we focus on this \q~range, since the intensities of the $3d_{xy}$ and $3d_{xz}$ excitations are here stronger than for negative \q, thus allowing to better resolve their contributions). {\bf e} ({\bf f}), Comparison between experiment (blue dotted line) and single ion theory (green solid line) for \q=1.0 (\q=0.5), respectively.}
\label{fig:SMdd}
\end{figure*}

For the case of purely local orbital excitations, the theoretical RIXS spectrum at a specific \q~ can be constructed as the sum of all the orbital contributions considered as Gaussian line shapes with momentum independent width $\Gamma_i$ and energy position $E_i$ [see Fig.\,S1(b)]. While the intensity of each Gaussian curve is given by the single ion responses at a specific geometry \q~ as discussed in \cite{Moretti2011}, the momentum-independent width of the nondispersive excitations and the on-site energy of the orbital excitations are free parameters of this model, which we determine from the comparison with the experiment while keeping the orbital sequence obtained by quantum chemical calculations \cite{Huang2011}. For the case of \cco, there is a very good agreement between the local model and the experiment at \q=0 [see Fig.\,S1(c) and (d)], using the 
Gaussian parameters 
\kwf{shown in}
Table \ref{t1}.
 
\begin{table}[h!]
\begin{center}
\begin{tabular}{|l|c|c|c|c|}
\hline
 \cco  & {3$d_{xy}$} & {3$d_{yz}$} & {3$d_{xz}$}  & {3$d_{3z^2-r^2}$} \\
\hline
{\bf E(eV)} & 1.62 & 1.84 & 2.03 & 2.35 \\
{\bf $\Gamma$ (eV)} & 0.18 & 0.18 & 0.19 & 0.36 \\
\hline  
\end{tabular}
\caption{Gaussian fitting parameters of orbital excitations in \cco, in the 
\kwf{local model}
($3d_{xz}$ parameters optimized at \q=0).}
\label{t1}
\end{center}
\end{table}
However, for \q $\neq$0 important deviations are observed, as can be clearly seen in Figure S1(e) and (f). 
These clear deviations from the local model are due to momentum dependent $E_i$ and $\Gamma_{i}$ of the 3$d_{xz}$ orbital excitation. The comparison of the experimental data to the local model therefore, on the on hand, confirms the local nature of the 3$d_{xy}$, 3$d_{yz}$  and 3$d_{3z^2-r^2}$ orbital excitations; on the other hand it raises a question about the dispersive nature of the 3$d_{xz}$ excitation, which is the central focus of the paper.

It is demonstrated in Figure 3(d) in the main text that the asymmetric and broad 3$d_{xz}$ experimental lineshape is perfectly accounted for by the orbiton spectral function calculated using Equation (1) and shown in Figure 3(b) (both referring to the main text; cf. Sec. {\bf C} below). The calculated spectral function not only presents a momentum dependent line shape / width, but also has a momentum dependent peak position with a maximum dispersion of $\sim$50 meV. 
In order to perform a direct comparison between the theoretical spectral function and the experimental RIXS data, the spectral function has to be multiplied by the single ion response function, which accounts for the specific geometry of the actual experiment. As can be observed in Figure 3(d) in the main text (red line), the resulting theoretical RIXS spectrum is now in extremely good agreement with the experimental data, in the whole range of momentum \q~ that is accessible in RIXS.

At the same time, the very good agreement between the 3$d_{xy}$ orbital excitation and the local picture strongly confirms the confinement of the 3$d_{xy}$ orbital excitation (cf. \kwf{Sec.}\,{\bf D}).

\section{C. Spin-orbital superexchange model for the 3$d_{xz}$ orbital excitation}

As mentioned in the main text of the paper the dominant orbital superexchange processes for the 3$d_{xz}$ orbital excitation are along the \kw{ladder} leg direction $x$ and follow from the relatively large hopping between the 3$d_{xz}$ orbital and the neighboring oxygen  2$p_z$ orbital, [cf. Fig. S2(a) and Fig. 1(c) in the main text].  Taking into account not only all possible superexchange processes contributing to the superexchange along this direction (which take place both on copper and on oxygen sites, cf. \cite{Schlappa2012, Wohlfeld2013}) 
but also the full multiplet structure of the virtual doubly occupied oxygen 2$p^2$
and copper 3$d^2$ configuration \cite{Schlappa2012, Wohlfeld2013} we obtain 
\begin{align}
 J^{xz}_{\rm {leg}}\! =\! \frac{(2 t^{\pi 1}_{\rm {leg}} t^{\sigma}_{\rm {leg}})^2  R_{xz}}{ U (\Delta_\sigma+V_{dp})(\Delta_\pi+V_{dp})} \simeq 0.051\ {\rm eV},
\end{align}
where (in the hole notation, cf. \cite{Schlappa2012}, and presenting only the absolute values of the hoppings below)
$t^{\pi 1}_{\rm {leg}} \simeq 0.61$ ($t^{\sigma}_{\rm {leg}} \simeq 1.16$) eV is the hopping element between Cu 3$d_{xz}$ and O 2$p_z$ orbital 
($d_{x^2-y^2}$ and $p_x$ orbital), $U \simeq 8.8$ eV is the Hubbard repulsion on Cu site, 
$\Delta_\sigma \simeq 3.0 $ eV ($\Delta_\pi \simeq 1.5$ eV) is the on-site charge transfer energy from 3$d_{x^2-y^2}$ (3$d_{xz}$ or 3$d_{xy}$) 
orbital to one of the oxygen 2$p$ orbitals, $V_{dp} \simeq 1$ eV is the intersite Coulomb 
repulsion. Finally, 
$R_{xz}$ takes care of the multiplet structure of the virtual excited states on copper and oxygen and reads
\begin{align}
 R_{\rm xz} = (3 R^b_1 + 3r^b_1 + R^b_2+ r^b_2 )/8 \simeq 2.08,
\end{align}
where $r^b_1 = 1/(1-3J^b_H/U)$, $r^b_2 = 1/(1-J^b_H/U)$,
$R^b_1 = 2U/[\Delta_\sigma+\Delta_\pi+U_p(1-3J^p_H/U_p)]$,
and $R^b_2 = 2U/[\Delta_\sigma+\Delta_\pi+U_p(1-J^p_H/U_p)]$.
Except for the hopping elements obtained from the band structure FPLO calculation \cite{Koepernik1999} the parameters follow from 
Ref. \cite{Neudert2000} and we assumed for the Hubbard repulsion on the oxygen site $U_p \simeq 4.4 $ eV, 
for the Hund's exchange for electrons in the $a\equiv 3d_{x^2-y^2}$ and $b\equiv 3d_{xz}$ orbitals $ J^b_H \equiv J^{ab}_H \simeq 1.2 $ eV,  
for the Hund's exchange for electrons in the $a\equiv 3d_{x^2-y^2}$ and $c\equiv 3d_{xy}$ orbitals
$J^c_H \equiv J^{ac}_H \simeq 0.69$ eV, and  
for the Hund's exchange for electrons on the oxygen site $J^p_H \simeq 0.83$ eV. 

Next, we look at the possible superexchange processes along the rung direction $y$ of the ladder, cf. Figures S2(a) and S2(c).
It occurs that $J^{xz}_{\rm {rung}} =0$ because the hopping element from the copper 3$d_{xz}$ orbital to the weakly 
hybdridizing oxygen 2$p_x$ orbital 
along the rung direction $y$ is non zero only due to the buckling of the ladders but nevertheless it is negligible 
(it is of the order of $\simeq 0.04$ eV in our FPLO calculations).

As mentioned in the main text the 90$^\circ$ Cu-O-Cu angle between the ladders leads to very weak and frustrated interladder spin superexchange which can be neglected. {\it A prori},
also the orbital superexchange should not be affected by the presence of the neighboring ladder. While this is indeed true in the ionic picture, this changes
if we take into account strong covalency effects present in \cco~(Note that: (i) quantum chemical calculations \cite{HozoiPrivComm} show that \cco~ is even more covalent 
than \sco~ for which it was already reported that covalency effects played a huge role in e.g. its magnetic properties \cite{Walters2009}, and 
(ii) e.g. Ref. \cite{Kim2003} suggests that in the ground state there are as many holes on the oxygen as on the copper.) Thus the strong 
hybridization between the neighboring ladders ($t^{\sigma}_{\rm il} \simeq 1.22$ eV) leads to a finite number of holes to be located in 
the 2$p_y$ orbital on oxygen and this renormalizes the superexchange $J^{xz}_{\rm {leg}}$ due to the stronger Coulomb repulsion on oxygen 
($U_p$) when this hole is present. In our qualitative model for the 3$d_{xz}$ orbital, however,  we can skip this interaction; instead a similar interladder covalency effect will become crucial for the analysis of the 3$d_{xy}$ orbital excitation.

Altogether, combining the orbital superexchange interaction derived above with the well-known result for the spin superexchange
interaction in the buckled ladder \cco, we obtain the following Hamiltonian which jointly describes the spin and orbital
superexchange
\begin{align} \label{eq:tJ1sup}
{\mathcal {{H}}} \equiv  {\mathcal {{H}}_{xz}}	 = &   - J^{xz}_{\rm {leg}} \!\sum_{\langle {\bf i}, {\bf j} \rangle || x , \sigma} 
({o}^\dag_{{\bf i} \sigma, {xz}}
{o}_{{\bf j} \sigma, {xz}}\! +\! h.c.) 
\kwf{+E_0^{xz} \sum_{\bf i} n_{{\bf i}, xz}}
\nonumber \\ 
 &+ J_{\rm {rung}} \sum_{\langle {\bf i}, {\bf j} \rangle || y } {\bf S}_{{\bf i} } \cdot {\bf S}_{\bf j} 
 + J_{\rm leg } \sum_{\langle {\bf i}, {\bf j} \rangle || x } {\bf S}_{{\bf i} } \cdot {\bf S}_{{\bf j}},
\end{align}
i.e. Equation (1) in the main text of the paper \kwf{(the notation used in the above equation
is also explained in the main text of the paper)}. 
Throughout the paper we assume the values of the spin superexchange
following Ref. \cite{Bordas2005} which gives $J_{\rm {leg}} \simeq 0.134$ eV,
and $J_{\rm {rung}} \simeq 0.011$ eV and agrees well with experimental
considerations from Ref. \cite{Lake2010}. Let us also stress that
both the experimental value of $J_{\rm {leg}} \simeq 160$ meV as decuced
from our RIXS experiment as well as the spin superexchange
as calculated using the hopping elements from
our FPLO calculations and used to obtain the values of
the orbital superexchange agrees within 15\% accuracy
with this commonly accepted value of $J_{\rm {leg}}$. Note that the
notation for the orbital superexchange (explained in the
main text of the paper) is different than the notation for
the spin superexchange as the above spin-orbital model
is written in the `$t$-$J$' model representation \cite{Wohlfeld2011} which is
very convenient for studying orbiton propagation.
\kwf{Finally, let us note that in the above equation, besides the superexchange interactions,
we also included the on-site energy cost of creating an orbital excitation in the 3$d_{xz}$ orbital.
While the energy cost of such local orbital excitation can be well-estimated using the 
quantum chemical calculations \cite{Huang2011} ($E_0^{xz}=\vbf{1.87}$ eV), in this paper we deduce 
this number directly from the RIXS experiment and obtain $E_0^{xz}=\vbf{2.03}$ eV (which actually matches
very well the fully {\it ab-initio} and parameter free quantum chemical calculations).}


\section{D. Spin-orbital superexchange model for the 3$d_{xy}$ orbital excitation}

We now turn our attention to the possible propagation of the 3$d_{xy}$ orbital excitation. A similar analysis as above {\it at first sight} leads 
to two finite paths for orbital superexchange: along the ladder rung and leg,  cf. Figures S2(b) and S2(c).
This is because from the 3$d_{xy}$ orbital the hole can hop along both the $x$ and $y$ direction, despite the buckling of the ladder.
Thus we obtain
\begin{align}
 J^{xy}_{\rm {leg}} = \frac{(2 t^{\pi 2}_{\rm {leg}} t^{\sigma}_{\rm {leg}})^2  R_{xy}}{ U (\Delta_\sigma+V_{dp})(\Delta_\pi+V_{dp})} \simeq 0.042 \ {\rm eV},
\end{align}
where $t^{\pi 2}_{\rm {leg}} \simeq 0.58$ eV is the hopping element between Cu $d_{xy}$ and O $p_y$ orbital and 
\begin{align}
 R_{\rm xy} = (3 R^c_1 + 3r^c_1 + R^c_2+ r^c_2 )/8 \simeq 1.93,
\end{align}
takes care of the multiplet structure (see Sec. C) where
$r^c_1 = 1/{1-3J^c_H/U}$,  $r^c_2 = 1/(1-J^c_H/U)$,
$R^c_1 = 2U/[\Delta_\sigma+\Delta_\pi+U_p(1-3J^p_H/U_p)]$,
and $R^c_2 = 2U/[\Delta_\sigma+\Delta_\pi+U_p(1-J^p_H/U_p)]$.

However, the hybridization with the neighboring ladder ($t^{\sigma}_{\rm il} \simeq 1.22$ eV), as discussed in the main text, is not negligible [cf. Fig. S2(b)]. This is because the presence of holes in the oxygen 2$p_y$ state cause the {\it blocking} of the superexchange processes along the leg for the 3$d_{xy}$ orbital: 
if the hole occupying this orbital has its spin pointing in the same direction as the hole in the 3$d_{xy}$ orbital (and due to frustration in spin
interaction between the ladders \cite{Lake2010}, there is 50\% probability for this situation), then the superexchange for 3$d_{xy}$ orbital is prohibited
[cf. Fig. S2(b)]. In this way the $J^{xy}_{\rm {leg}}$ orbital superexchange along the leg for the 3$d_{xy}$ orbital cannot lead to coherent 
motion of the orbiton along the ladder leg. In general, this shows another peculiar interplay between neighboring ladders which, also in many other circumstances, should not 
be considered as decoupled ~\cite{Wohlfeld2010}.

\begin{figure}[t!]
\center{
\includegraphics[width=0.72\columnwidth]{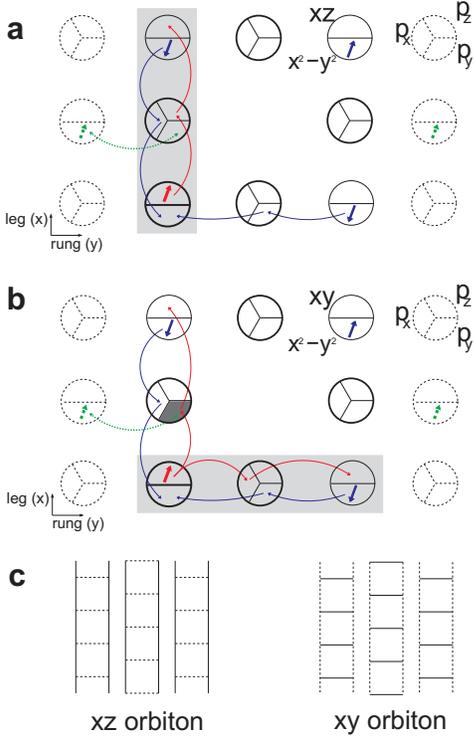}}
\caption{(a) [(b)] Schematic view of the orbital superexchange processes for the 3$d_{xz}$ (3$d_{xy}$) orbital excitation: the copper (oxygen) sites are depicted with circles consisting of two (three) orbitals relevant for the particular superexchange processes which can only be finite if {\it two} single-headed arrows (indicating the hopping between orbitals) connect the copper sites, respectively. The double-headed arrows show possible delocalization of the 3$d_{x^2-y^2}$ hole from the neighboring ladder into the oxygen orbital in the ladder under consideration which may lead to blocking of superexchange processes due to Pauli principle [cf. the 2$p_y$ orbital colored in black in panel (b)] -- thus the grey areas show these Cu-O-Cu bonds along which the orbital superexchange is finite. (c) Finite orbital superexchange processes as obtained from panel (a) [(b)] for 3$d_{xz}$ (3$d_{xy}$) orbiton represented as solid lines in the ladder geometry (the dotted lines show bonds along which only the spin superexchange can be finite). 
}
\label{fig:cartoon_superexchange}
\end{figure}

On the other hand, the orbital superexchange along the rung is not affected by the above processes [cf. Fig.\, S2(b)] 
and is equal to 
\begin{align}
 J^{xy}_{\rm {rung}} = \frac{4 (t^{\pi}_{\rm {rung}})^2 [(t^{\sigma}_{\rm {rung}})^2 - (t^{{\sigma \pi}}_{\rm {rung}})^2]  R_{xy}}{U (\Delta_\sigma+V_{dp})(\Delta_\pi+V_{dp})} \simeq 0.074  \ {\rm eV},
\end{align}
where $t^{\pi}_{\rm {rung}} \simeq 0.81 $ eV ($t^{\sigma}_{\rm {rung}} \simeq 1.10$ eV) is the hopping element between Cu 3$d_{xy}$ and O 2$p_x$ orbital 
(3$d_{x^2-y^2}$ and 2$p_y$ orbital), and $t^{{\sigma \pi}}_{\rm {rung}} \simeq 0.56$ eV is the hopping element between 3$d_{x^2-y^2}$ and 2$p_z$ orbital 
which is finite due to the buckling of the ladders. Note that the minus sign in $(t^{\sigma}_{\rm {rung}})^2 - (t^{{\sigma \pi}}_{\rm {rung}})^2 $ in the above equation follows from the different signs of the
hopping elements between the 2$p_z$ orbital and the 3$d_{x^2-y^2}$ to the left / right in the rung.

Adding the anisotropic spin superexchange interactions $J_{\rm {leg}}$ and $J_{\rm {rung}}$ \kwf{(and the on-site
energy of the orbital excitation $E_0^{xy}=\vbf{1.84}$ eV as obtained from the RIXS experiment, cf. Sec. {\bf B})}, 
we obtain:
\begin{align} \label{eq:tJ2}
{\mathcal {{H}}}^{xy} =&  - J^{xy}_{\rm {rung}}\sum_{\langle {\bf i}, {\bf j} \rangle || y , \sigma} ({o}^\dag_{{\bf i} \sigma, {xy}}
{o}_{{\bf j} \sigma, {xy}}\! +\! h.c.)   
\kwf{+E_0^{xy} \sum_{\bf i} n_{{\bf i}, xy}}
\nonumber \\ 
 &+ J_{\rm {rung}} \sum_{\langle {\bf i}, {\bf j} \rangle || y } {\bf S}_{{\bf i} } \cdot {\bf S}_{\bf j} 
 + J_{\rm {leg}} \sum_{\langle {\bf i}, {\bf j} \rangle || x } {\bf S}_{{\bf i} } \cdot {\bf S}_{{\bf j}}.
\end{align}
Again $J_{\rm {rung}} \ll J_{\rm {leg}}$, but since the 3$d_{xy}$ orbiton cannot hop along the leg now (see above),
it is confined along the rung direction $x$ and cannot lead to the orbital dispersion along the leg $y$ direction [cf. Fig.\, S2(c)].

\section*{E. `Robustness' of spin-orbital separation.}

As discussed in the main text, in general the spin-orbital separation phenomenon seems to be very robust because: (i) the motion of the 3$d_{xz}$ orbital excitation is strictly 1D, and (ii) spinons are much faster than orbitons (due to $J_{\rm {leg}} \gg J^{xz}_{\rm {leg}}$) which enables an easy separation of the two.

The ``degree of fractionalization'' introduced in the main text allows to compare the degree of spin-orbital (hole) separation between different systems. Such a quantity is the ratio between the correlation of a spin at r=0 with an orbital (hole) at r=$\infty$, and the correlation between a spin at r=0 with an orbital (hole) at r=1. If spin and orbital (hole) are fractionalized (separated) perfectly, then they are unperceiving to each other. Thus, the existence probability for a spin at r=0 is constant and independent from an orbital (hole) at a  position between r = 1:$\infty$. The ``degree of fractionalization'' should be therefore 1 in this case. When spin and orbital (hole) are bounded or mixed for some reason, the correlation instead has a maximum value at r=1 in the t-J model. As a result, the ratio is reduced from 1, which means that the ``degree of fractionalization'' is weakened. 

In this work, we performed finite-size-scaling analysis of the spin-orbital (hole) correlation to estimate its value at infinite distance. The results are represented in Fig. 4 of the main text  for the spin-orbital (hole) separation in \sco~ and \cco. For 2D systems, the degree of fractionalization is estimated to be zero since the correlation decays algebraically towards zero at infinite distance.

It occurs, firstly, that the spin-orbital separation in the buckled ladder \cco~ is almost as strong as the 
spin-orbital separation in \sco~ and the difference between them almost entirely stems from the weaker
spin superexchange along the ladder leg in \cco, cf. Fig. 4 in the main text. This striking similarity between these two cases
originates from the 1D (`directional') nature of the hopping of orbitals (cf. Fig. S2 and Fig. 1 from the main text):
had the hopping from the excited orbital been less anisotropic (e.g. 2D), 
the spin-orbital fractionalization would have been more or less destroyed (cf. the lack
of spin-charge fractionalization in the isotropic ladder or 2D cases). 
Secondly, {\it despite} the weak interladder spin interactions, the degree of spin-orbital separation in \cco~
is much {\it larger} than the degree of fractionalization of the spin and charge in \sco, cf. Fig. 4 in the main text.
This happens because the ratio of the velocities of spinon and orbiton / holon is entirely different
in these two cases: in \sco~ spinon is slower than holon ($J<t$ in a typical $t$--$J$ model describing spin-charge separation 
\cite{Kim1996, Fujisawa1999, Kim2006}), while in \cco~ spinon is faster than orbiton [cf. $J_{\rm {leg}} \gg J^{xz}_{\rm {leg}}$ in 
Eq. (S3) and Ref. \cite{Wohlfeld2011}]. Therefore, the spinon can move away much quicker from the orbiton
than from the holon, i.e. from the quasiparticle at which we look in the experiment allowing to verify the degree of separation from the spinon. 

Finally, let us here comment on the question whether this generic `robustness' of the spin-orbital separation with respect to spin-charge separation
might mean that the latter cannot be seen in \cco~ (as stated above, so far it has not been observed experimentally). Here, we suggest, 
that the answer depends on the values of the effective hopping parameters in \cco: while according to \cite{Bordas2005} the only other (than the leg
direction) non-negligible hopping is the rung hopping $t$ which is three times smaller than the leg hopping and allows for a degree of fractionalization 
substantial enough to be observed (it is ca. 10\% smaller than in \sco), we cannot exclude that in reality other long-range
hoppings are still large enough to destroy the fractionalization. This is because, the hopping from the 3$d_{x^2-y^2}$ is not `directional'
and the ratio of $J/t$ is low, leading to a weak fractionalization regime already in the case of spin-charge separation in \sco.


\begin{thebibliography}{38}%
\makeatletter
\providecommand \@ifxundefined [1]{%
 \@ifx{#1\undefined}
}%
\providecommand \@ifnum [1]{%
 \ifnum #1\expandafter \@firstoftwo
 \else \expandafter \@secondoftwo
 \fi
}%
\providecommand \@ifx [1]{%
 \ifx #1\expandafter \@firstoftwo
 \else \expandafter \@secondoftwo
 \fi
}%
\providecommand \natexlab [1]{#1}%
\providecommand \enquote  [1]{``#1''}%
\providecommand \bibnamefont  [1]{#1}%
\providecommand \bibfnamefont [1]{#1}%
\providecommand \citenamefont [1]{#1}%
\providecommand \href@noop [0]{\@secondoftwo}%
\providecommand \href [0]{\begingroup \@sanitize@url \@href}%
\providecommand \@href[1]{\@@startlink{#1}\@@href}%
\providecommand \@@href[1]{\endgroup#1\@@endlink}%
\providecommand \@sanitize@url [0]{\catcode `\\12\catcode `\$12\catcode
  `\&12\catcode `\#12\catcode `\^12\catcode `\_12\catcode `\%12\relax}%
\providecommand \@@startlink[1]{}%
\providecommand \@@endlink[0]{}%
\providecommand \url  [0]{\begingroup\@sanitize@url \@url }%
\providecommand \@url [1]{\endgroup\@href {#1}{\urlprefix }}%
\providecommand \urlprefix  [0]{URL }%
\providecommand \Eprint [0]{\href }%
\providecommand \doibase [0]{http://dx.doi.org/}%
\providecommand \selectlanguage [0]{\@gobble}%
\providecommand \bibinfo  [0]{\@secondoftwo}%
\providecommand \bibfield  [0]{\@secondoftwo}%
\providecommand \translation [1]{[#1]}%
\providecommand \BibitemOpen [0]{}%
\providecommand \bibitemStop [0]{}%
\providecommand \bibitemNoStop [0]{.\EOS\space}%
\providecommand \EOS [0]{\spacefactor3000\relax}%
\providecommand \BibitemShut  [1]{\csname bibitem#1\endcsname}%
\let\auto@bib@innerbib\@empty
\bibitem [{\citenamefont {Voit}(1995)}]{Voit1995}%
  \BibitemOpen
  \bibfield  {author} {\bibinfo {author} {\bibfnamefont {J.}~\bibnamefont
  {Voit}},\ }\href@noop {} {\bibfield  {journal} {\bibinfo  {journal} {Reports
  on Progress in Physics}\ }\textbf {\bibinfo {volume} {58}},\ \bibinfo {pages}
  {977} (\bibinfo {year} {1995})}\BibitemShut {NoStop}%
\bibitem [{\citenamefont {Kim}\ \emph {et~al.}(1996)\citenamefont {Kim},
  \citenamefont {Matsuura}, \citenamefont {Shen}, \citenamefont {Motoyama},
  \citenamefont {Eisaki}, \citenamefont {Uchida}, \citenamefont {Tohyama},\
  and\ \citenamefont {Maekawa}}]{Kim1996}%
  \BibitemOpen
  \bibfield  {author} {\bibinfo {author} {\bibfnamefont {C.}~\bibnamefont
  {Kim}}, \bibinfo {author} {\bibfnamefont {A.~Y.}\ \bibnamefont {Matsuura}},
  \bibinfo {author} {\bibfnamefont {Z.-X.}\ \bibnamefont {Shen}}, \bibinfo
  {author} {\bibfnamefont {N.}~\bibnamefont {Motoyama}}, \bibinfo {author}
  {\bibfnamefont {H.}~\bibnamefont {Eisaki}}, \bibinfo {author} {\bibfnamefont
  {S.}~\bibnamefont {Uchida}}, \bibinfo {author} {\bibfnamefont
  {T.}~\bibnamefont {Tohyama}}, \ and\ \bibinfo {author} {\bibfnamefont
  {S.}~\bibnamefont {Maekawa}},\ }\href {\doibase 10.1103/PhysRevLett.77.4054}
  {\bibfield  {journal} {\bibinfo  {journal} {Phys. Rev. Lett.}\ }\textbf
  {\bibinfo {volume} {77}},\ \bibinfo {pages} {4054} (\bibinfo {year}
  {1996})}\BibitemShut {NoStop}%
\bibitem [{\citenamefont {Wohlfeld}\ \emph {et~al.}(2011)\citenamefont
  {Wohlfeld}, \citenamefont {Daghofer}, \citenamefont {Nishimoto},
  \citenamefont {Khaliullin},\ and\ \citenamefont {van~den
  Brink}}]{Wohlfeld2011}%
  \BibitemOpen
  \bibfield  {author} {\bibinfo {author} {\bibfnamefont {K.}~\bibnamefont
  {Wohlfeld}}, \bibinfo {author} {\bibfnamefont {M.}~\bibnamefont {Daghofer}},
  \bibinfo {author} {\bibfnamefont {S.}~\bibnamefont {Nishimoto}}, \bibinfo
  {author} {\bibfnamefont {G.}~\bibnamefont {Khaliullin}}, \ and\ \bibinfo
  {author} {\bibfnamefont {J.}~\bibnamefont {van~den Brink}},\ }\href {\doibase
  10.1103/PhysRevLett.107.147201} {\bibfield  {journal} {\bibinfo  {journal}
  {Physical Review Letters}\ }\textbf {\bibinfo {volume} {107}},\ \bibinfo
  {pages} {147201} (\bibinfo {year} {2011})}\BibitemShut {NoStop}%
\bibitem [{\citenamefont {Schlappa}\ \emph {et~al.}(2012)\citenamefont
  {Schlappa}, \citenamefont {Wohlfeld}, \citenamefont {Zhou}, \citenamefont
  {Mourigal}, \citenamefont {Haverkort}, \citenamefont {Strocov}, \citenamefont
  {Hozoi}, \citenamefont {Monney}, \citenamefont {Nishimoto}, \citenamefont
  {Singh}, \citenamefont {Revcolevschi}, \citenamefont {Caux}, \citenamefont
  {Patthey}, \citenamefont {Rønnow}, \citenamefont {Brink},\ and\
  \citenamefont {Schmitt}}]{Schlappa2012}%
  \BibitemOpen
  \bibfield  {author} {\bibinfo {author} {\bibfnamefont {J.}~\bibnamefont
  {Schlappa}}, \bibinfo {author} {\bibfnamefont {K.}~\bibnamefont {Wohlfeld}},
  \bibinfo {author} {\bibfnamefont {K.~J.}\ \bibnamefont {Zhou}}, \bibinfo
  {author} {\bibfnamefont {M.}~\bibnamefont {Mourigal}}, \bibinfo {author}
  {\bibfnamefont {M.~W.}\ \bibnamefont {Haverkort}}, \bibinfo {author}
  {\bibfnamefont {V.~N.}\ \bibnamefont {Strocov}}, \bibinfo {author}
  {\bibfnamefont {L.}~\bibnamefont {Hozoi}}, \bibinfo {author} {\bibfnamefont
  {C.}~\bibnamefont {Monney}}, \bibinfo {author} {\bibfnamefont
  {S.}~\bibnamefont {Nishimoto}}, \bibinfo {author} {\bibfnamefont
  {S.}~\bibnamefont {Singh}}, \bibinfo {author} {\bibfnamefont
  {A.}~\bibnamefont {Revcolevschi}}, \bibinfo {author} {\bibfnamefont
  {J.}~\bibnamefont {Caux}}, \bibinfo {author} {\bibfnamefont {L.}~\bibnamefont
  {Patthey}}, \bibinfo {author} {\bibfnamefont {H.~M.}\ \bibnamefont
  {Rønnow}}, \bibinfo {author} {\bibfnamefont {J.~v.~d.}\ \bibnamefont
  {Brink}}, \ and\ \bibinfo {author} {\bibfnamefont {T.}~\bibnamefont
  {Schmitt}},\ }\href {\doibase 10.1038/nature10974} {\bibfield  {journal}
  {\bibinfo  {journal} {Nature}\ }\textbf {\bibinfo {volume} {485}},\ \bibinfo
  {pages} {82} (\bibinfo {year} {2012})}\BibitemShut {NoStop}%
\bibitem [{\citenamefont {Moreno}\ \emph {et~al.}(2013)\citenamefont {Moreno},
  \citenamefont {Muramatsu},\ and\ \citenamefont {Carmelo}}]{Moreno2013}%
  \BibitemOpen
  \bibfield  {author} {\bibinfo {author} {\bibfnamefont {A.}~\bibnamefont
  {Moreno}}, \bibinfo {author} {\bibfnamefont {A.}~\bibnamefont {Muramatsu}}, \
  and\ \bibinfo {author} {\bibfnamefont {J.~M.~P.}\ \bibnamefont {Carmelo}},\
  }\href {\doibase 10.1103/PhysRevB.87.075101} {\bibfield  {journal} {\bibinfo
  {journal} {Phys. Rev. B}\ }\textbf {\bibinfo {volume} {87}},\ \bibinfo
  {pages} {075101} (\bibinfo {year} {2013})}\BibitemShut {NoStop}%
\bibitem [{\citenamefont {Anderson}(1987)}]{Anderson1987}%
  \BibitemOpen
  \bibfield  {author} {\bibinfo {author} {\bibfnamefont {P.~W.}\ \bibnamefont
  {Anderson}},\ }\href@noop {} {\bibfield  {journal} {\bibinfo  {journal}
  {Science}\ }\textbf {\bibinfo {volume} {235}},\ \bibinfo {pages} {1196}
  (\bibinfo {year} {1987})}\BibitemShut {NoStop}%
\bibitem [{\citenamefont {Senthil}\ and\ \citenamefont
  {Fisher}(2001)}]{Senthil2001}%
  \BibitemOpen
  \bibfield  {author} {\bibinfo {author} {\bibfnamefont {T.}~\bibnamefont
  {Senthil}}\ and\ \bibinfo {author} {\bibfnamefont {M.~P.~A.}\ \bibnamefont
  {Fisher}},\ }\href {\doibase 10.1103/PhysRevLett.86.292} {\bibfield
  {journal} {\bibinfo  {journal} {Phys. Rev. Lett.}\ }\textbf {\bibinfo
  {volume} {86}},\ \bibinfo {pages} {292} (\bibinfo {year} {2001})}\BibitemShut
  {NoStop}%
\bibitem [{\citenamefont {Holt}\ \emph {et~al.}(2012)\citenamefont {Holt},
  \citenamefont {Oitmaa}, \citenamefont {Chen},\ and\ \citenamefont
  {Sushkov}}]{Holt2012}%
  \BibitemOpen
  \bibfield  {author} {\bibinfo {author} {\bibfnamefont {M.}~\bibnamefont
  {Holt}}, \bibinfo {author} {\bibfnamefont {J.}~\bibnamefont {Oitmaa}},
  \bibinfo {author} {\bibfnamefont {W.}~\bibnamefont {Chen}}, \ and\ \bibinfo
  {author} {\bibfnamefont {O.~P.}\ \bibnamefont {Sushkov}},\ }\href {\doibase
  10.1103/PhysRevLett.109.037001} {\bibfield  {journal} {\bibinfo  {journal}
  {Phys. Rev. Lett.}\ }\textbf {\bibinfo {volume} {109}},\ \bibinfo {pages}
  {037001} (\bibinfo {year} {2012})}\BibitemShut {NoStop}%
\bibitem [{nho()}]{nhole}%
  \BibitemOpen
  \href@noop {} {}\bibinfo {note} {In the introductory paragraph and in the
  caption of Fig.\,1, the term {\it hole} refers to the hole formed in the
  $S$=1/2 spin background, i.e. the removal of the electron with its spin from
  the 3$d_{x^2-y^2}$ orbital. Note that in the rest of the paper this term
  refers to the hole as defined by `hole notation' widely used to describe the
  copper oxides: in that notation the one hole in the ground state
  3$d_{x^2-y^2}$ orbital arises from the particle-hole transformation to the
  3$d$ shell filled with 9 electrons (with 3$d_{x^2-y^2}$ orbital being
  half-filled).}\BibitemShut {Stop}%
\bibitem [{\citenamefont {Martinez}\ and\ \citenamefont
  {Horsch}(1991)}]{Martinez1991}%
  \BibitemOpen
  \bibfield  {author} {\bibinfo {author} {\bibfnamefont {G.}~\bibnamefont
  {Martinez}}\ and\ \bibinfo {author} {\bibfnamefont {P.}~\bibnamefont
  {Horsch}},\ }\href {\doibase 10.1103/PhysRevB.44.317} {\bibfield  {journal}
  {\bibinfo  {journal} {Phys.~Rev.~B}\ }\textbf {\bibinfo {volume} {44}},\
  \bibinfo {pages} {317} (\bibinfo {year} {1991})}\BibitemShut {NoStop}%
\bibitem [{\citenamefont {Troyer}\ \emph {et~al.}(1996)\citenamefont {Troyer},
  \citenamefont {Tsunetsugu},\ and\ \citenamefont {Rice}}]{Troyer1996}%
  \BibitemOpen
  \bibfield  {author} {\bibinfo {author} {\bibfnamefont {M.}~\bibnamefont
  {Troyer}}, \bibinfo {author} {\bibfnamefont {H.}~\bibnamefont {Tsunetsugu}},
  \ and\ \bibinfo {author} {\bibfnamefont {T.~M.}\ \bibnamefont {Rice}},\
  }\href {\doibase 10.1103/PhysRevB.53.251} {\bibfield  {journal} {\bibinfo
  {journal} {Phys. Rev. B}\ }\textbf {\bibinfo {volume} {53}},\ \bibinfo
  {pages} {251} (\bibinfo {year} {1996})}\BibitemShut {NoStop}%
\bibitem [{\citenamefont {White}\ and\ \citenamefont
  {Scalapino}(1997)}]{White1997}%
  \BibitemOpen
  \bibfield  {author} {\bibinfo {author} {\bibfnamefont {S.~R.}\ \bibnamefont
  {White}}\ and\ \bibinfo {author} {\bibfnamefont {D.~J.}\ \bibnamefont
  {Scalapino}},\ }\href {\doibase 10.1103/PhysRevB.55.6504} {\bibfield
  {journal} {\bibinfo  {journal} {Phys. Rev. B}\ }\textbf {\bibinfo {volume}
  {55}},\ \bibinfo {pages} {6504} (\bibinfo {year} {1997})}\BibitemShut
  {NoStop}%
\bibitem [{nla()}]{nladder}%
  \BibitemOpen
  \href@noop {} {}\bibinfo {note} {Note that the Luttinger liquid theory, which
  predicts spin-charge separation in the 1D case, does not apply to the
  $t$--$J$ model on the ladder with gapped spin excitations
  \textcolor{blue}{\b[13]}.}\BibitemShut {Stop}%
\bibitem [{\citenamefont {{Dagotto}}\ and\ \citenamefont
  {{Rice}}(1996)}]{Dagotto1996}%
  \BibitemOpen
  \bibfield  {author} {\bibinfo {author} {\bibfnamefont {E.}~\bibnamefont
  {{Dagotto}}}\ and\ \bibinfo {author} {\bibfnamefont {T.~M.}\ \bibnamefont
  {{Rice}}},\ }\href {\doibase 10.1126/science.271.5249.618} {\bibfield
  {journal} {\bibinfo  {journal} {Science}\ }\textbf {\bibinfo {volume}
  {271}},\ \bibinfo {pages} {618} (\bibinfo {year} {1996})}\BibitemShut
  {NoStop}%
\bibitem [{\citenamefont {Ghiringhelli}\ \emph {et~al.}(2004)\citenamefont
  {Ghiringhelli}, \citenamefont {Brookes}, \citenamefont {Annese},
  \citenamefont {Berger}, \citenamefont {Dallera}, \citenamefont {Grioni},
  \citenamefont {Perfetti}, \citenamefont {Tagliaferri},\ and\ \citenamefont
  {Braicovich}}]{Ghiringhelli2004}%
  \BibitemOpen
  \bibfield  {author} {\bibinfo {author} {\bibfnamefont {G.}~\bibnamefont
  {Ghiringhelli}}, \bibinfo {author} {\bibfnamefont {N.~B.}\ \bibnamefont
  {Brookes}}, \bibinfo {author} {\bibfnamefont {E.}~\bibnamefont {Annese}},
  \bibinfo {author} {\bibfnamefont {H.}~\bibnamefont {Berger}}, \bibinfo
  {author} {\bibfnamefont {C.}~\bibnamefont {Dallera}}, \bibinfo {author}
  {\bibfnamefont {M.}~\bibnamefont {Grioni}}, \bibinfo {author} {\bibfnamefont
  {L.}~\bibnamefont {Perfetti}}, \bibinfo {author} {\bibfnamefont
  {A.}~\bibnamefont {Tagliaferri}}, \ and\ \bibinfo {author} {\bibfnamefont
  {L.}~\bibnamefont {Braicovich}},\ }\href {\doibase
  10.1103/PhysRevLett.92.117406} {\bibfield  {journal} {\bibinfo  {journal}
  {Phys. Rev. Lett.}\ }\textbf {\bibinfo {volume} {92}},\ \bibinfo {pages}
  {117406} (\bibinfo {year} {2004})}\BibitemShut {NoStop}%
\bibitem [{\citenamefont {Braicovich}\ \emph {et~al.}(2009)\citenamefont
  {Braicovich}, \citenamefont {Ament}, \citenamefont {Bisogni}, \citenamefont
  {Forte}, \citenamefont {Aruta}, \citenamefont {Balestrino}, \citenamefont
  {Brookes}, \citenamefont {{De Luca}}, \citenamefont {Medaglia}, \citenamefont
  {Granozio}, \citenamefont {Radovic}, \citenamefont {Salluzzo}, \citenamefont
  {van~den Brink},\ and\ \citenamefont {Ghiringhelli}}]{Braicovich2009}%
  \BibitemOpen
  \bibfield  {author} {\bibinfo {author} {\bibfnamefont {L.}~\bibnamefont
  {Braicovich}}, \bibinfo {author} {\bibfnamefont {L.~J.~P.}\ \bibnamefont
  {Ament}}, \bibinfo {author} {\bibfnamefont {V.}~\bibnamefont {Bisogni}},
  \bibinfo {author} {\bibfnamefont {F.}~\bibnamefont {Forte}}, \bibinfo
  {author} {\bibfnamefont {C.}~\bibnamefont {Aruta}}, \bibinfo {author}
  {\bibfnamefont {G.}~\bibnamefont {Balestrino}}, \bibinfo {author}
  {\bibfnamefont {N.~B.}\ \bibnamefont {Brookes}}, \bibinfo {author}
  {\bibfnamefont {G.~M.}\ \bibnamefont {{De Luca}}}, \bibinfo {author}
  {\bibfnamefont {P.~G.}\ \bibnamefont {Medaglia}}, \bibinfo {author}
  {\bibfnamefont {F.~M.}\ \bibnamefont {Granozio}}, \bibinfo {author}
  {\bibfnamefont {M.}~\bibnamefont {Radovic}}, \bibinfo {author} {\bibfnamefont
  {M.}~\bibnamefont {Salluzzo}}, \bibinfo {author} {\bibfnamefont
  {J.}~\bibnamefont {van~den Brink}}, \ and\ \bibinfo {author} {\bibfnamefont
  {G.}~\bibnamefont {Ghiringhelli}},\ }\href {\doibase
  10.1103/PhysRevLett.102.167401} {\bibfield  {journal} {\bibinfo  {journal}
  {Physical Review Letters}\ }\textbf {\bibinfo {volume} {102}},\ \bibinfo
  {pages} {167401} (\bibinfo {year} {2009})}\BibitemShut {NoStop}%
\bibitem [{\citenamefont {Schlappa}\ \emph {et~al.}(2009)\citenamefont
  {Schlappa}, \citenamefont {Schmitt}, \citenamefont {Vernay}, \citenamefont
  {Strocov}, \citenamefont {Ilakovac}, \citenamefont {Thielemann},
  \citenamefont {R\o{}nnow}, \citenamefont {Vanishri}, \citenamefont
  {Piazzalunga}, \citenamefont {Wang}, \citenamefont {Braicovich},
  \citenamefont {Ghiringhelli}, \citenamefont {Marin}, \citenamefont {Mesot},
  \citenamefont {Delley},\ and\ \citenamefont {Patthey}}]{Schlappa2009}%
  \BibitemOpen
  \bibfield  {author} {\bibinfo {author} {\bibfnamefont {J.}~\bibnamefont
  {Schlappa}}, \bibinfo {author} {\bibfnamefont {T.}~\bibnamefont {Schmitt}},
  \bibinfo {author} {\bibfnamefont {F.}~\bibnamefont {Vernay}}, \bibinfo
  {author} {\bibfnamefont {V.~N.}\ \bibnamefont {Strocov}}, \bibinfo {author}
  {\bibfnamefont {V.}~\bibnamefont {Ilakovac}}, \bibinfo {author}
  {\bibfnamefont {B.}~\bibnamefont {Thielemann}}, \bibinfo {author}
  {\bibfnamefont {H.~M.}\ \bibnamefont {R\o{}nnow}}, \bibinfo {author}
  {\bibfnamefont {S.}~\bibnamefont {Vanishri}}, \bibinfo {author}
  {\bibfnamefont {A.}~\bibnamefont {Piazzalunga}}, \bibinfo {author}
  {\bibfnamefont {X.}~\bibnamefont {Wang}}, \bibinfo {author} {\bibfnamefont
  {L.}~\bibnamefont {Braicovich}}, \bibinfo {author} {\bibfnamefont
  {G.}~\bibnamefont {Ghiringhelli}}, \bibinfo {author} {\bibfnamefont
  {C.}~\bibnamefont {Marin}}, \bibinfo {author} {\bibfnamefont
  {J.}~\bibnamefont {Mesot}}, \bibinfo {author} {\bibfnamefont
  {B.}~\bibnamefont {Delley}}, \ and\ \bibinfo {author} {\bibfnamefont
  {L.}~\bibnamefont {Patthey}},\ }\href {\doibase
  10.1103/PhysRevLett.103.047401} {\bibfield  {journal} {\bibinfo  {journal}
  {Physical Review Letters}\ }\textbf {\bibinfo {volume} {103}},\ \bibinfo
  {pages} {047401} (\bibinfo {year} {2009})}\BibitemShut {NoStop}%
\bibitem [{\citenamefont {Ament}\ \emph {et~al.}(2011)\citenamefont {Ament},
  \citenamefont {van Veenendaal}, \citenamefont {Devereaux}, \citenamefont
  {Hill},\ and\ \citenamefont {van~den Brink}}]{Ament2011}%
  \BibitemOpen
  \bibfield  {author} {\bibinfo {author} {\bibfnamefont {L.~J.~P.}\
  \bibnamefont {Ament}}, \bibinfo {author} {\bibfnamefont {M.}~\bibnamefont
  {van Veenendaal}}, \bibinfo {author} {\bibfnamefont {T.~P.}\ \bibnamefont
  {Devereaux}}, \bibinfo {author} {\bibfnamefont {J.~P.}\ \bibnamefont {Hill}},
  \ and\ \bibinfo {author} {\bibfnamefont {J.}~\bibnamefont {van~den Brink}},\
  }\href {\doibase 10.1103/RevModPhys.83.705} {\bibfield  {journal} {\bibinfo
  {journal} {Review of Modern Physics}\ }\textbf {\bibinfo {volume} {83}},\
  \bibinfo {pages} {705} (\bibinfo {year} {2011})}\BibitemShut {NoStop}%
\bibitem [{\citenamefont {Bisogni}\ \emph {et~al.}(2014)\citenamefont
  {Bisogni}, \citenamefont {Kourtis}, \citenamefont {Monney}, \citenamefont
  {Zhou}, \citenamefont {Kraus}, \citenamefont {Sekar}, \citenamefont
  {Strocov}, \citenamefont {Buechner}, \citenamefont {Brink}, \citenamefont
  {Braicovich}, \citenamefont {Schmitt}, \citenamefont {Daghofer},\ and\
  \citenamefont {Geck}}]{Bisogni2013}%
  \BibitemOpen
  \bibfield  {author} {\bibinfo {author} {\bibfnamefont {V.}~\bibnamefont
  {Bisogni}}, \bibinfo {author} {\bibfnamefont {S.}~\bibnamefont {Kourtis}},
  \bibinfo {author} {\bibfnamefont {C.}~\bibnamefont {Monney}}, \bibinfo
  {author} {\bibfnamefont {K.}~\bibnamefont {Zhou}}, \bibinfo {author}
  {\bibfnamefont {R.}~\bibnamefont {Kraus}}, \bibinfo {author} {\bibfnamefont
  {C.}~\bibnamefont {Sekar}}, \bibinfo {author} {\bibfnamefont
  {V.}~\bibnamefont {Strocov}}, \bibinfo {author} {\bibfnamefont
  {B.}~\bibnamefont {Buechner}}, \bibinfo {author} {\bibfnamefont {J.~v.~d.}\
  \bibnamefont {Brink}}, \bibinfo {author} {\bibfnamefont {L.}~\bibnamefont
  {Braicovich}}, \bibinfo {author} {\bibfnamefont {T.}~\bibnamefont {Schmitt}},
  \bibinfo {author} {\bibfnamefont {M.}~\bibnamefont {Daghofer}}, \ and\
  \bibinfo {author} {\bibfnamefont {J.}~\bibnamefont {Geck}},\ }\href {\doibase
  10.1103/PhysRevLett.112.147401} {\bibfield  {journal} {\bibinfo  {journal}
  {Phys. Rev. Letters}\ }\textbf {\bibinfo {volume} {112}},\ \bibinfo {pages}
  {147401} (\bibinfo {year} {2014})}\BibitemShut {NoStop}%
\bibitem [{SM()}]{SM}%
  \BibitemOpen
  \href@noop {} {}\bibinfo {note} {See Supplementary Material [url], which includes Refs. [31-38], for details about the experimental configuration, the data analysis and the spin-orbital superexchange model for the $3d_{xz}$ and $3d_{xy}$ orbital.}\BibitemShut {Stop}%
\bibitem [{\citenamefont {Lake}\ \emph {et~al.}(2010)\citenamefont {Lake},
  \citenamefont {Tsvelik}, \citenamefont {Notbohm}, \citenamefont
  {Alan~Tennant}, \citenamefont {Perring}, \citenamefont {Reehuis},
  \citenamefont {Sekar}, \citenamefont {Krabbes},\ and\ \citenamefont
  {Buchner}}]{Lake2010}%
  \BibitemOpen
  \bibfield  {author} {\bibinfo {author} {\bibfnamefont {B.}~\bibnamefont
  {Lake}}, \bibinfo {author} {\bibfnamefont {A.~M.}\ \bibnamefont {Tsvelik}},
  \bibinfo {author} {\bibfnamefont {S.}~\bibnamefont {Notbohm}}, \bibinfo
  {author} {\bibfnamefont {D.}~\bibnamefont {Alan~Tennant}}, \bibinfo {author}
  {\bibfnamefont {T.~G.}\ \bibnamefont {Perring}}, \bibinfo {author}
  {\bibfnamefont {M.}~\bibnamefont {Reehuis}}, \bibinfo {author} {\bibfnamefont
  {C.}~\bibnamefont {Sekar}}, \bibinfo {author} {\bibfnamefont
  {G.}~\bibnamefont {Krabbes}}, \ and\ \bibinfo {author} {\bibfnamefont
  {B.}~\bibnamefont {Buchner}},\ }\href {\doibase 10.1038/nphys1462} {\bibfield
   {journal} {\bibinfo  {journal} {Nature Physics}\ }\textbf {\bibinfo {volume}
  {6}},\ \bibinfo {pages} {50} (\bibinfo {year} {2010})}\BibitemShut {NoStop}%
\bibitem [{\citenamefont {Bordas}\ \emph {et~al.}(2005)\citenamefont {Bordas},
  \citenamefont {de~Graaf}, \citenamefont {Caballol},\ and\ \citenamefont
  {Calzado}}]{Bordas2005}%
  \BibitemOpen
  \bibfield  {author} {\bibinfo {author} {\bibfnamefont {E.}~\bibnamefont
  {Bordas}}, \bibinfo {author} {\bibfnamefont {C.}~\bibnamefont {de~Graaf}},
  \bibinfo {author} {\bibfnamefont {R.}~\bibnamefont {Caballol}}, \ and\
  \bibinfo {author} {\bibfnamefont {C.~J.}\ \bibnamefont {Calzado}},\ }\href
  {\doibase 10.1103/PhysRevB.71.045108} {\bibfield  {journal} {\bibinfo
  {journal} {Phys. Rev. B}\ }\textbf {\bibinfo {volume} {71}},\ \bibinfo
  {pages} {045108} (\bibinfo {year} {2005})}\BibitemShut {NoStop}%
\bibitem [{\citenamefont {Ghiringhelli}\ \emph {et~al.}(2006)\citenamefont
  {Ghiringhelli}, \citenamefont {Piazzalunga}, \citenamefont {Dallera},
  \citenamefont {Trezzi}, \citenamefont {Braicovich}, \citenamefont {Schmitt},
  \citenamefont {Strocov}, \citenamefont {Betemps}, \citenamefont {Patthey},
  \citenamefont {Wang},\ and\ \citenamefont {Grioni}}]{saxes}%
  \BibitemOpen
  \bibfield  {author} {\bibinfo {author} {\bibfnamefont {G.}~\bibnamefont
  {Ghiringhelli}}, \bibinfo {author} {\bibfnamefont {A.}~\bibnamefont
  {Piazzalunga}}, \bibinfo {author} {\bibfnamefont {C.}~\bibnamefont
  {Dallera}}, \bibinfo {author} {\bibfnamefont {G.}~\bibnamefont {Trezzi}},
  \bibinfo {author} {\bibfnamefont {L.}~\bibnamefont {Braicovich}}, \bibinfo
  {author} {\bibfnamefont {T.}~\bibnamefont {Schmitt}}, \bibinfo {author}
  {\bibfnamefont {V.~N.}\ \bibnamefont {Strocov}}, \bibinfo {author}
  {\bibfnamefont {R.}~\bibnamefont {Betemps}}, \bibinfo {author} {\bibfnamefont
  {L.}~\bibnamefont {Patthey}}, \bibinfo {author} {\bibfnamefont
  {X.}~\bibnamefont {Wang}}, \ and\ \bibinfo {author} {\bibfnamefont
  {M.}~\bibnamefont {Grioni}},\ }\href@noop {} {\bibfield  {journal} {\bibinfo
  {journal} {Review of Scientific Instruments}\ }\textbf {\bibinfo {volume}
  {77}},\ \bibinfo {pages} {113108} (\bibinfo {year} {2006})}\BibitemShut
  {NoStop}%
\bibitem [{\citenamefont {Strocov}\ \emph {et~al.}(2010)\citenamefont
  {Strocov}, \citenamefont {Schmitt}, \citenamefont {Flechsig}, \citenamefont
  {Schmidt}, \citenamefont {Imhof}, \citenamefont {Chen}, \citenamefont
  {Raabe}, \citenamefont {Betemps}, \citenamefont {Zimoch}, \citenamefont
  {Krempasky}, \citenamefont {Piazzalunga}, \citenamefont {Wang}, \citenamefont
  {Grioni},\ and\ \citenamefont {Patthey}}]{Strocov2010}%
  \BibitemOpen
  \bibfield  {author} {\bibinfo {author} {\bibfnamefont {V.~N.}\ \bibnamefont
  {Strocov}}, \bibinfo {author} {\bibfnamefont {T.}~\bibnamefont {Schmitt}},
  \bibinfo {author} {\bibfnamefont {U.}~\bibnamefont {Flechsig}}, \bibinfo
  {author} {\bibfnamefont {T.}~\bibnamefont {Schmidt}}, \bibinfo {author}
  {\bibfnamefont {A.}~\bibnamefont {Imhof}}, \bibinfo {author} {\bibfnamefont
  {Q.}~\bibnamefont {Chen}}, \bibinfo {author} {\bibfnamefont {J.}~\bibnamefont
  {Raabe}}, \bibinfo {author} {\bibfnamefont {R.}~\bibnamefont {Betemps}},
  \bibinfo {author} {\bibfnamefont {D.}~\bibnamefont {Zimoch}}, \bibinfo
  {author} {\bibfnamefont {J.}~\bibnamefont {Krempasky}}, \bibinfo {author}
  {\bibfnamefont {A.}~\bibnamefont {Piazzalunga}}, \bibinfo {author}
  {\bibfnamefont {X.}~\bibnamefont {Wang}}, \bibinfo {author} {\bibfnamefont
  {M.}~\bibnamefont {Grioni}}, \ and\ \bibinfo {author} {\bibfnamefont
  {L.}~\bibnamefont {Patthey}},\ }\href@noop {} {\bibfield  {journal} {\bibinfo
   {journal} {J. Synchrotron Rad.}\ }\textbf {\bibinfo {volume} {17}},\
  \bibinfo {pages} {631} (\bibinfo {year} {2010})}\BibitemShut {NoStop}%
\bibitem [{\citenamefont {Huang}\ \emph {et~al.}(2011)\citenamefont {Huang},
  \citenamefont {Bogdanov}, \citenamefont {Siurakshina}, \citenamefont {Fulde},
  \citenamefont {van~den Brink},\ and\ \citenamefont {Hozoi}}]{Huang2011}%
  \BibitemOpen
  \bibfield  {author} {\bibinfo {author} {\bibfnamefont {H.-Y.}\ \bibnamefont
  {Huang}}, \bibinfo {author} {\bibfnamefont {N.~A.}\ \bibnamefont {Bogdanov}},
  \bibinfo {author} {\bibfnamefont {L.}~\bibnamefont {Siurakshina}}, \bibinfo
  {author} {\bibfnamefont {P.}~\bibnamefont {Fulde}}, \bibinfo {author}
  {\bibfnamefont {J.}~\bibnamefont {van~den Brink}}, \ and\ \bibinfo {author}
  {\bibfnamefont {L.}~\bibnamefont {Hozoi}},\ }\href {\doibase
  10.1103/PhysRevB.84.235125} {\bibfield  {journal} {\bibinfo  {journal}
  {Physical Review B}\ }\textbf {\bibinfo {volume} {84}},\ \bibinfo {pages}
  {235125} (\bibinfo {year} {2011})}\BibitemShut {NoStop}%
\bibitem [{\citenamefont {Moretti~Sala}\ \emph {et~al.}(2011)\citenamefont
  {Moretti~Sala}, \citenamefont {Bisogni}, \citenamefont {Aruta}, \citenamefont
  {Balestrino}, \citenamefont {Berger}, \citenamefont {Brookes}, \citenamefont
  {{De Luca}}, \citenamefont {Di~Castro}, \citenamefont {Grioni}, \citenamefont
  {Guarise}, \citenamefont {Medaglia}, \citenamefont {Miletto~Granozio},
  \citenamefont {Minola}, \citenamefont {Perna}, \citenamefont {Radovic},
  \citenamefont {Salluzzo}, \citenamefont {Schmitt}, \citenamefont {Zhou},
  \citenamefont {Braicovich},\ and\ \citenamefont
  {Ghiringhelli}}]{Moretti2011}%
  \BibitemOpen
  \bibfield  {author} {\bibinfo {author} {\bibfnamefont {M.}~\bibnamefont
  {Moretti~Sala}}, \bibinfo {author} {\bibfnamefont {V.}~\bibnamefont
  {Bisogni}}, \bibinfo {author} {\bibfnamefont {C.}~\bibnamefont {Aruta}},
  \bibinfo {author} {\bibfnamefont {G.}~\bibnamefont {Balestrino}}, \bibinfo
  {author} {\bibfnamefont {H.}~\bibnamefont {Berger}}, \bibinfo {author}
  {\bibfnamefont {N.~B.}\ \bibnamefont {Brookes}}, \bibinfo {author}
  {\bibfnamefont {G.~M.}\ \bibnamefont {{De Luca}}}, \bibinfo {author}
  {\bibfnamefont {D.}~\bibnamefont {Di~Castro}}, \bibinfo {author}
  {\bibfnamefont {M.}~\bibnamefont {Grioni}}, \bibinfo {author} {\bibfnamefont
  {M.}~\bibnamefont {Guarise}}, \bibinfo {author} {\bibfnamefont {P.~G.}\
  \bibnamefont {Medaglia}}, \bibinfo {author} {\bibfnamefont {F.}~\bibnamefont
  {Miletto~Granozio}}, \bibinfo {author} {\bibfnamefont {M.}~\bibnamefont
  {Minola}}, \bibinfo {author} {\bibfnamefont {P.}~\bibnamefont {Perna}},
  \bibinfo {author} {\bibfnamefont {M.}~\bibnamefont {Radovic}}, \bibinfo
  {author} {\bibfnamefont {M.}~\bibnamefont {Salluzzo}}, \bibinfo {author}
  {\bibfnamefont {T.}~\bibnamefont {Schmitt}}, \bibinfo {author} {\bibfnamefont
  {K.~J.}\ \bibnamefont {Zhou}}, \bibinfo {author} {\bibfnamefont
  {L.}~\bibnamefont {Braicovich}}, \ and\ \bibinfo {author} {\bibfnamefont
  {G.}~\bibnamefont {Ghiringhelli}},\ }\href {\doibase
  10.1088/1367-2630/13/4/043026} {\bibfield  {journal} {\bibinfo  {journal}
  {New Journal of Physics}\ }\textbf {\bibinfo {volume} {13}},\ \bibinfo
  {pages} {043026} (\bibinfo {year} {2011})}\BibitemShut {NoStop}%
\bibitem [{\citenamefont {Koepernik}\ and\ \citenamefont
  {Eschrig}(1999)}]{Koepernik1999}%
  \BibitemOpen
  \bibfield  {author} {\bibinfo {author} {\bibfnamefont {K.}~\bibnamefont
  {Koepernik}}\ and\ \bibinfo {author} {\bibfnamefont {H.}~\bibnamefont
  {Eschrig}},\ }\href {\doibase 10.1103/PhysRevB.59.1743} {\bibfield  {journal}
  {\bibinfo  {journal} {Phys. Rev. B}\ }\textbf {\bibinfo {volume} {59}},\
  \bibinfo {pages} {1743} (\bibinfo {year} {1999})}\BibitemShut {NoStop}%
\bibitem [{\citenamefont {Wohlfeld}\ \emph {et~al.}(2013)\citenamefont
  {Wohlfeld}, \citenamefont {Nishimoto}, \citenamefont {Haverkort},\ and\
  \citenamefont {van~den Brink}}]{Wohlfeld2013}%
  \BibitemOpen
  \bibfield  {author} {\bibinfo {author} {\bibfnamefont {K.}~\bibnamefont
  {Wohlfeld}}, \bibinfo {author} {\bibfnamefont {S.}~\bibnamefont {Nishimoto}},
  \bibinfo {author} {\bibfnamefont {M.~W.}\ \bibnamefont {Haverkort}}, \ and\
  \bibinfo {author} {\bibfnamefont {J.}~\bibnamefont {van~den Brink}},\ }\href
  {\doibase 10.1103/PhysRevB.88.195138} {\bibfield  {journal} {\bibinfo
  {journal} {Phys. Rev. B}\ }\textbf {\bibinfo {volume} {88}},\ \bibinfo
  {pages} {195138} (\bibinfo {year} {2013})}\BibitemShut {NoStop}%
\bibitem [{\citenamefont {Harris}\ \emph {et~al.}(2003)\citenamefont {Harris},
  \citenamefont {Yildirim}, \citenamefont {Aharony}, \citenamefont
  {Entin-Wohlman},\ and\ \citenamefont {Korenblit}}]{Harris2003}%
  \BibitemOpen
  \bibfield  {author} {\bibinfo {author} {\bibfnamefont {A.~B.}\ \bibnamefont
  {Harris}}, \bibinfo {author} {\bibfnamefont {T.}~\bibnamefont {Yildirim}},
  \bibinfo {author} {\bibfnamefont {A.}~\bibnamefont {Aharony}}, \bibinfo
  {author} {\bibfnamefont {O.}~\bibnamefont {Entin-Wohlman}}, \ and\ \bibinfo
  {author} {\bibfnamefont {I.~Y.}\ \bibnamefont {Korenblit}},\ }\href {\doibase
  10.1103/PhysRevLett.91.087206} {\bibfield  {journal} {\bibinfo  {journal}
  {Phys. Rev. Lett.}\ }\textbf {\bibinfo {volume} {91}},\ \bibinfo {pages}
  {087206} (\bibinfo {year} {2003})}\BibitemShut {NoStop}%
\bibitem [{\citenamefont {Wohlfeld}\ \emph {et~al.}(2008)\citenamefont
  {Wohlfeld}, \citenamefont {Daghofer}, \citenamefont
  {Ole\ifmmode~\acute{s}\else \'{s}\fi{}},\ and\ \citenamefont
  {Horsch}}]{Wohlfeld2008}%
  \BibitemOpen
  \bibfield  {author} {\bibinfo {author} {\bibfnamefont {K.}~\bibnamefont
  {Wohlfeld}}, \bibinfo {author} {\bibfnamefont {M.}~\bibnamefont {Daghofer}},
  \bibinfo {author} {\bibfnamefont {A.~M.}\ \bibnamefont
  {Ole\ifmmode~\acute{s}\else \'{s}\fi{}}}, \ and\ \bibinfo {author}
  {\bibfnamefont {P.}~\bibnamefont {Horsch}},\ }\href {\doibase
  10.1103/PhysRevB.78.214423} {\bibfield  {journal} {\bibinfo  {journal} {Phys.
  Rev. B}\ }\textbf {\bibinfo {volume} {78}},\ \bibinfo {pages} {214423}
  (\bibinfo {year} {2008})}\BibitemShut {NoStop}%
\bibitem [{\citenamefont {Sekar}\ \emph {et~al.}(2005)\citenamefont {Sekar},
  \citenamefont {Krabbes},\ and\ \citenamefont {Teresiak}}]{Sekar2005}%
  \BibitemOpen
  \bibfield  {author} {\bibinfo {author} {\bibfnamefont {C.}~\bibnamefont
  {Sekar}}, \bibinfo {author} {\bibfnamefont {G.}~\bibnamefont {Krabbes}}, \
  and\ \bibinfo {author} {\bibfnamefont {A.}~\bibnamefont {Teresiak}},\
  }\href@noop {} {\bibfield  {journal} {\bibinfo  {journal} {Journal of Crystal
  Growth}\ }\textbf {\bibinfo {volume} {273}},\ \bibinfo {pages} {403}
  (\bibinfo {year} {2005})}\BibitemShut {NoStop}%
\bibitem [{\citenamefont {Neudert}\ \emph {et~al.}(2000)\citenamefont
  {Neudert}, \citenamefont {Drechsler}, \citenamefont {M\'alek}, \citenamefont
  {Rosner}, \citenamefont {Kielwein}, \citenamefont {Hu}, \citenamefont
  {Knupfer}, \citenamefont {Golden}, \citenamefont {Fink}, \citenamefont
  {N\"ucker}, \citenamefont {Merz}, \citenamefont {Schuppler}, \citenamefont
  {Motoyama}, \citenamefont {Eisaki}, \citenamefont {Uchida}, \citenamefont
  {Domke},\ and\ \citenamefont {Kaindl}}]{Neudert2000}%
  \BibitemOpen
  \bibfield  {author} {\bibinfo {author} {\bibfnamefont {R.}~\bibnamefont
  {Neudert}}, \bibinfo {author} {\bibfnamefont {S.-L.}\ \bibnamefont
  {Drechsler}}, \bibinfo {author} {\bibfnamefont {J.}~\bibnamefont {M\'alek}},
  \bibinfo {author} {\bibfnamefont {H.}~\bibnamefont {Rosner}}, \bibinfo
  {author} {\bibfnamefont {M.}~\bibnamefont {Kielwein}}, \bibinfo {author}
  {\bibfnamefont {Z.}~\bibnamefont {Hu}}, \bibinfo {author} {\bibfnamefont
  {M.}~\bibnamefont {Knupfer}}, \bibinfo {author} {\bibfnamefont {M.~S.}\
  \bibnamefont {Golden}}, \bibinfo {author} {\bibfnamefont {J.}~\bibnamefont
  {Fink}}, \bibinfo {author} {\bibfnamefont {N.}~\bibnamefont {N\"ucker}},
  \bibinfo {author} {\bibfnamefont {M.}~\bibnamefont {Merz}}, \bibinfo {author}
  {\bibfnamefont {S.}~\bibnamefont {Schuppler}}, \bibinfo {author}
  {\bibfnamefont {N.}~\bibnamefont {Motoyama}}, \bibinfo {author}
  {\bibfnamefont {H.}~\bibnamefont {Eisaki}}, \bibinfo {author} {\bibfnamefont
  {S.}~\bibnamefont {Uchida}}, \bibinfo {author} {\bibfnamefont
  {M.}~\bibnamefont {Domke}}, \ and\ \bibinfo {author} {\bibfnamefont
  {G.}~\bibnamefont {Kaindl}},\ }\href {\doibase 10.1103/PhysRevB.62.10752}
  {\bibfield  {journal} {\bibinfo  {journal} {Phys. Rev. B}\ }\textbf {\bibinfo
  {volume} {62}},\ \bibinfo {pages} {10752} (\bibinfo {year}
  {2000})}\BibitemShut {NoStop}%
\bibitem [{Hoz()}]{HozoiPrivComm}%
  \BibitemOpen
  \href@noop {} {}\bibinfo {note} {L. Hozoi {\it et al. priv.
  comm.}}\BibitemShut {Stop}%
\bibitem [{\citenamefont {Walters}\ \emph {et~al.}(2009)\citenamefont
  {Walters}, \citenamefont {Perring}, \citenamefont {Caux}, \citenamefont
  {Savici}, \citenamefont {Gu}, \citenamefont {Lee}, \citenamefont {Ku},\ and\
  \citenamefont {Zaliznyak}}]{Walters2009}%
  \BibitemOpen
  \bibfield  {author} {\bibinfo {author} {\bibfnamefont {A.~C.}\ \bibnamefont
  {Walters}}, \bibinfo {author} {\bibfnamefont {T.~G.}\ \bibnamefont
  {Perring}}, \bibinfo {author} {\bibfnamefont {J.}~\bibnamefont {Caux}},
  \bibinfo {author} {\bibfnamefont {A.~T.}\ \bibnamefont {Savici}}, \bibinfo
  {author} {\bibfnamefont {G.~D.}\ \bibnamefont {Gu}}, \bibinfo {author}
  {\bibfnamefont {C.}~\bibnamefont {Lee}}, \bibinfo {author} {\bibfnamefont
  {W.}~\bibnamefont {Ku}}, \ and\ \bibinfo {author} {\bibfnamefont {I.~A.}\
  \bibnamefont {Zaliznyak}},\ }\href {\doibase 10.1038/nphys1405} {\bibfield
  {journal} {\bibinfo  {journal} {Nature Physics}\ }\textbf {\bibinfo {volume}
  {5}},\ \bibinfo {pages} {867} (\bibinfo {year} {2009})}\BibitemShut {NoStop}%
\bibitem [{\citenamefont {Kim}\ \emph {et~al.}(2003)\citenamefont {Kim},
  \citenamefont {Rosner}, \citenamefont {Drechsler}, \citenamefont {Hu},
  \citenamefont {Sekar}, \citenamefont {Krabbes}, \citenamefont {M\'alek},
  \citenamefont {Knupfer}, \citenamefont {Fink},\ and\ \citenamefont
  {Eschrig}}]{Kim2003}%
  \BibitemOpen
  \bibfield  {author} {\bibinfo {author} {\bibfnamefont {T.~K.}\ \bibnamefont
  {Kim}}, \bibinfo {author} {\bibfnamefont {H.}~\bibnamefont {Rosner}},
  \bibinfo {author} {\bibfnamefont {S.-L.}\ \bibnamefont {Drechsler}}, \bibinfo
  {author} {\bibfnamefont {Z.}~\bibnamefont {Hu}}, \bibinfo {author}
  {\bibfnamefont {C.}~\bibnamefont {Sekar}}, \bibinfo {author} {\bibfnamefont
  {G.}~\bibnamefont {Krabbes}}, \bibinfo {author} {\bibfnamefont
  {J.}~\bibnamefont {M\'alek}}, \bibinfo {author} {\bibfnamefont
  {M.}~\bibnamefont {Knupfer}}, \bibinfo {author} {\bibfnamefont
  {J.}~\bibnamefont {Fink}}, \ and\ \bibinfo {author} {\bibfnamefont
  {H.}~\bibnamefont {Eschrig}},\ }\href {\doibase 10.1103/PhysRevB.67.024516}
  {\bibfield  {journal} {\bibinfo  {journal} {Physical Review B}\ }\textbf
  {\bibinfo {volume} {67}},\ \bibinfo {pages} {024516} (\bibinfo {year}
  {2003})}\BibitemShut {NoStop}%
\bibitem [{\citenamefont {Wohlfeld}\ \emph {et~al.}(2010)\citenamefont
  {Wohlfeld}, \citenamefont {Ole\ifmmode~\acute{s}\else \'{s}\fi{}},\ and\
  \citenamefont {Sawatzky}}]{Wohlfeld2010}%
  \BibitemOpen
  \bibfield  {author} {\bibinfo {author} {\bibfnamefont {K.}~\bibnamefont
  {Wohlfeld}}, \bibinfo {author} {\bibfnamefont {A.~M.}\ \bibnamefont
  {Ole\ifmmode~\acute{s}\else \'{s}\fi{}}}, \ and\ \bibinfo {author}
  {\bibfnamefont {G.~A.}\ \bibnamefont {Sawatzky}},\ }\href {\doibase
  10.1103/PhysRevB.81.214522} {\bibfield  {journal} {\bibinfo  {journal} {Phys.
  Rev. B}\ }\textbf {\bibinfo {volume} {81}},\ \bibinfo {pages} {214522}
  (\bibinfo {year} {2010})}\BibitemShut {NoStop}%
\bibitem [{\citenamefont {Fujisawa}\ \emph {et~al.}(1999)\citenamefont
  {Fujisawa}, \citenamefont {Yokoya}, \citenamefont {Takahashi}, \citenamefont
  {Miyasaka}, \citenamefont {Kibune},\ and\ \citenamefont
  {Takagi}}]{Fujisawa1999}%
  \BibitemOpen
  \bibfield  {author} {\bibinfo {author} {\bibfnamefont {H.}~\bibnamefont
  {Fujisawa}}, \bibinfo {author} {\bibfnamefont {T.}~\bibnamefont {Yokoya}},
  \bibinfo {author} {\bibfnamefont {T.}~\bibnamefont {Takahashi}}, \bibinfo
  {author} {\bibfnamefont {S.}~\bibnamefont {Miyasaka}}, \bibinfo {author}
  {\bibfnamefont {M.}~\bibnamefont {Kibune}}, \ and\ \bibinfo {author}
  {\bibfnamefont {H.}~\bibnamefont {Takagi}},\ }\href {\doibase
  10.1103/PhysRevB.59.7358} {\bibfield  {journal} {\bibinfo  {journal} {Phys.
  Rev. B}\ }\textbf {\bibinfo {volume} {59}},\ \bibinfo {pages} {7358}
  (\bibinfo {year} {1999})}\BibitemShut {NoStop}%
\bibitem [{\citenamefont {Kim}\ \emph {et~al.}(2006)\citenamefont {Kim},
  \citenamefont {Koh}, \citenamefont {Rotenberg}, \citenamefont {Oh},
  \citenamefont {Eisaki}, \citenamefont {Motoyama}, \citenamefont {Uchida},
  \citenamefont {Tohyama}, \citenamefont {Maekawa},\ and\ \citenamefont
  {Shen}}]{Kim2006}%
  \BibitemOpen
  \bibfield  {author} {\bibinfo {author} {\bibfnamefont {B.~J.}\ \bibnamefont
  {Kim}}, \bibinfo {author} {\bibfnamefont {H.}~\bibnamefont {Koh}}, \bibinfo
  {author} {\bibfnamefont {E.}~\bibnamefont {Rotenberg}}, \bibinfo {author}
  {\bibfnamefont {S.-J.}\ \bibnamefont {Oh}}, \bibinfo {author} {\bibfnamefont
  {E.}~\bibnamefont {Eisaki}}, \bibinfo {author} {\bibfnamefont
  {N.}~\bibnamefont {Motoyama}}, \bibinfo {author} {\bibfnamefont
  {S.}~\bibnamefont {Uchida}}, \bibinfo {author} {\bibfnamefont
  {T.}~\bibnamefont {Tohyama}}, \bibinfo {author} {\bibfnamefont
  {S.}~\bibnamefont {Maekawa}}, \ and\ \bibinfo {author} {\bibfnamefont
  {C.}~\bibnamefont {Shen}, \bibfnamefont {Z.-X.and~Kim}},\ }\href {\doibase
  10.1038/nphys316} {\bibfield  {journal} {\bibinfo  {journal} {Nature
  Physics}\ }\textbf {\bibinfo {volume} {2}},\ \bibinfo {pages} {397} (\bibinfo
  {year} {2006})}\BibitemShut {NoStop}%
\end{thebibliography}
\end{document}